%% file: xi18_fin.tex
\documentclass[rmp,floatfix]{revtex4}

\usepackage{rotating,booktabs}
\usepackage{natbib}
\usepackage{times}
\usepackage{setspace}
\usepackage{amsmath, graphicx, amssymb,tabls}
\usepackage{graphicx}
\usepackage{bm}
\usepackage{amsmath, amsthm}

\usepackage{color}
\usepackage{enumerate}

\newcommand{\expt}[1]{\langle #1 \rangle}

\newcommand{\be}{\begin{equation}}
\newcommand{\ee}{\end{equation}}

\newcommand{\ra}{\rightarrow}

\usepackage[normalem]{ulem}

\begin{document}

\title{Demography-adjusted tests of neutrality based on genome-wide SNP data}

\author{M. Rafajlovi{\'c}$^{a,d}$, A. Klassmann$^{b,d}$, A. Eriksson$^{c}$, T. Wiehe$^{b}$,  and B. Mehlig$^{a}$\\
$^a$\emph{\small Department of Physics, University of Gothenburg, SE-41296 Gothenburg, Sweden}\\
$^b$\emph{\small Institut f{\"u}r Genetik, Universit{\"a}t zu K{\"o}ln, 50674 K{\"o}ln, Germany}\\
$^c$\emph{\small Department of Zoology, University of Cambridge, CB2 3EJ Cambridge, UK}\\
$^d$\emph{\small These authors have equally contributed to this work}\\
}

\begin{abstract}
Tests of the neutral evolution hypothesis are usually built on the standard null model which assumes that mutations are neutral and population size remains constant over time. 
However, it is unclear how such tests are affected if the last assumption is dropped. 
Here, we extend the unifying framework for tests based on the site frequency spectrum, introduced by Achaz and Ferretti, to populations of varying size.  
A key ingredient is to specify the first two moments of the frequency spectrum. 
We show that these moments can be determined analytically if a population has experienced two instantaneous size changes in the past. 
We apply our method to data from ten human populations gathered in the $1000$ genomes project,
estimate their demographies and define demography-adjusted versions of Tajima's $D$, Fay \& Wu's $H$, and Zeng's $E$. 
The adjusted test statistics facilitate the direct comparison between populations and they show that most of the differences among populations seen in the original tests can be explained by demography. We carried out
whole genome screens for deviation from neutrality and identified candidate regions of
recent positive selection. We provide track files with values of the adjusted and original tests
for upload to the UCSC genome browser.
\end{abstract}
\maketitle
\section{Introduction}
\begin{figure}[t]
\includegraphics[scale=0.8]{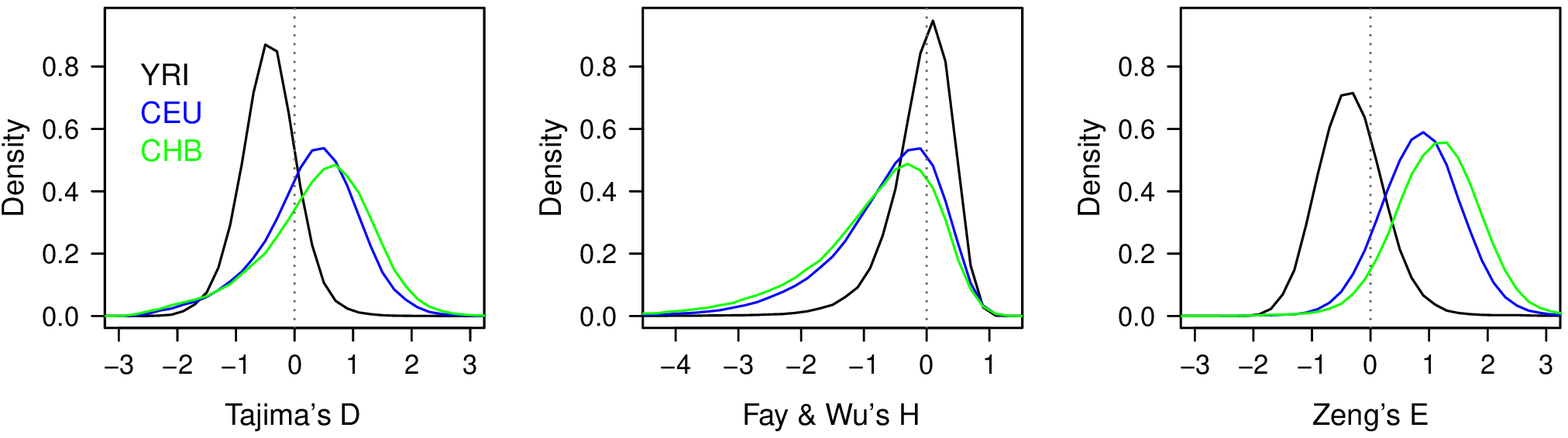}
\includegraphics[scale=0.8]{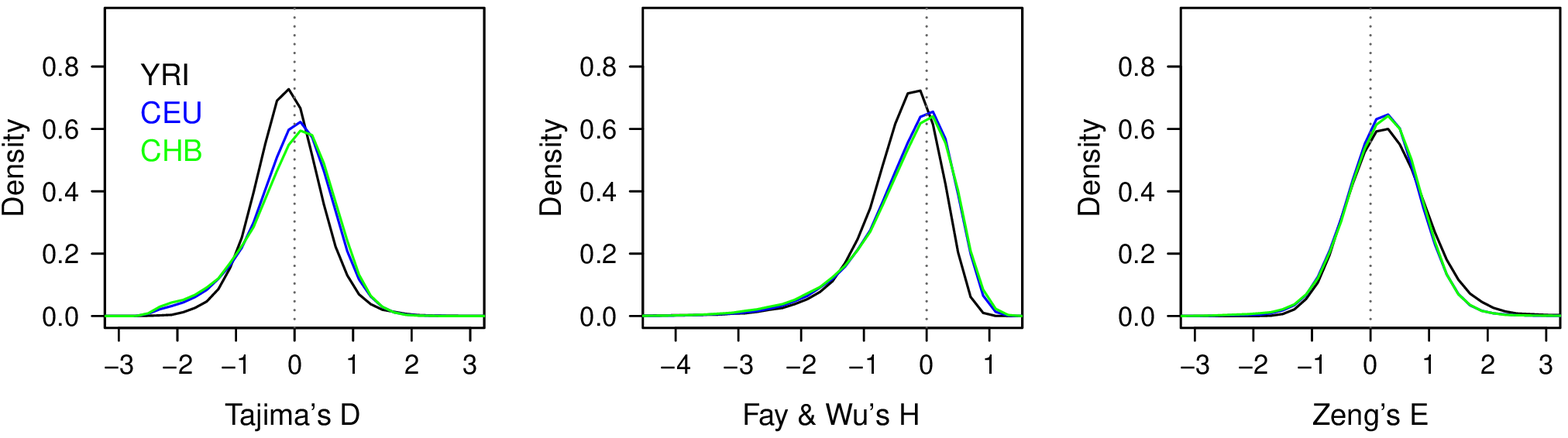}
\caption{\label{fig:densities}Distribution of test values over all sliding windows. Top row: original tests. 
Bottom row: demography-adjusted tests.}
\end{figure}

In natural populations, genetic diversity is shaped not only by population genetic forces such as drift and natural selection, but also by geographic structure and demographic history. Many statistical tests to identify genome regions affected by natural
selection have been proposed in the past, such as iHS \cite{iHS}, XP-EHH \cite{XP-EHH} as well as
Tajima's $D$ \cite{Taj:1989}, Fay\&Wu's $H$ \cite{FayWu:2000}, and Zengs's $E$ \cite{Zeng:2006}.
Tests of neutrality have frequently been used to search for signatures of selection in the human genome
\cite{Akey:2004, Stajich:2005,Carlson,Nielsen:human_chimp,iHS}. However, distinguishing selection from demographic effects in genomic data remains a challenge \cite{Akey:2004,Stajich:2005}. 
In this paper, we focus on tests based on the shape of the site frequency spectrum, such as  Tajima's $D$, Fay\&Wu's $H$, and  Zeng's $E$.
As examples, we show in Fig.~\ref{fig:densities} (upper panels) genome-wide values of these tests for a European (CEU),
Asian (CHB), and African human population (YRI) in the $1000$ genomes project dataset \cite{1000Genomes}. As
Fig.~\ref{fig:densities} (upper panels) shows, the distributions of the tests differ substantially between different
populations. To which extent do these differences arise from differences in demographic histories of the
populations? In order to answer this question, it is necessary to eliminate the effects of demographies on the values of
tests. In this study, we achieve this by adjusting the site frequency spectrum of tests of neutrality for the deviation
of population demographies from constant size. Thus, we modify tests of
neutrality by directly integrating demographies into them. We refer to such modified tests as {\it demography-adjusted}. When demography corresponds to constant population size, demography-adjusted tests reduce to the tests defined for the standard Wright-Fisher model, hereafter referred to as {\it original tests}.

The
distributions of demography-adjusted tests are similar to the distributions of the corresponding original tests computed
under the standard null model. Consequently, demography-adjusted tests significantly simplify a direct comparison of the
values of tests between different populations by emphasising the relevant differences. Examples are given in
Fig.~\ref{fig:densities} (lower panels), where we show the distributions of our demography-adjusted Tajima's $D$,
Fay\&Wu's $H$, and  Zeng's $E$ for the populations CEU, CHB, and YRI. As this figure suggests, most of the differences
in the distributions of the tests between human populations arise from their
distinct underlying demographies.

Since human demographies are unknown, it is necessary to estimate them. As suggested by
\citet{Nielsen:2000} (see also \citet{Adams}), we apply a maximum likelihood method to genome-wide single
nucleotide polymorphisms (SNPs). As an approximation for the demographies of human populations we use a simplified model with two instantaneous population size changes in the past, as proposed before \cite{Adams,Marth,Stajich:2005}. This model is characterized by four unknown parameters. It has the appealing property to yield exact analytical expressions for the first two moments of the site frequency spectrum (SFS). These are required to formulate our demography-adjusted tests of neutrality and they are explicitly derived in this paper.

The error in the estimate of demographic parameters 
depends on the noise in the genome-wide SFS, thus on the number of SNPs used for the estimation. We analyse the sensitivity of demography-adjusted tests by using coalescent simulations.  
On the basis of two reference demographies with two population-size changes in the past, we determine the number of SNPs required for reliable adjustment of the tests.
 
The populations CEU, CHB, and YRI are only three exemplary populations chosen from a set of ten populations analysed in this study by means of demography-adjusted tests. Assuming a piecewise constant demographic model, we find
 that Europeans and Asians  went through a recent population bottleneck,
which is in agreement with \citet{Adams} and \citet{Marth}. In
contrast, the African populations either experienced two population-size expansions (ASW, again in agreement with \citet{Adams} and \citet{Marth}), or an ancient expansion, followed by a recent population-size decline (LWK, YRI).

Our results further show that demography-adjustment of SFS-based tests is essentially reflected in an affine linear transformation of the test statistic. 
Consequently, the genomic regions recognized to be under selection by the adjusted tests 
strongly overlap with the originally detected regions. 
However, our adjusted tests permit a direct comparison of results from different 
populations with different demographies.

We provide original and adjusted tests values as BED-files, formatted for
upload to the UCSC genome browser.

\begin{figure}[t]
\centerline{
\includegraphics{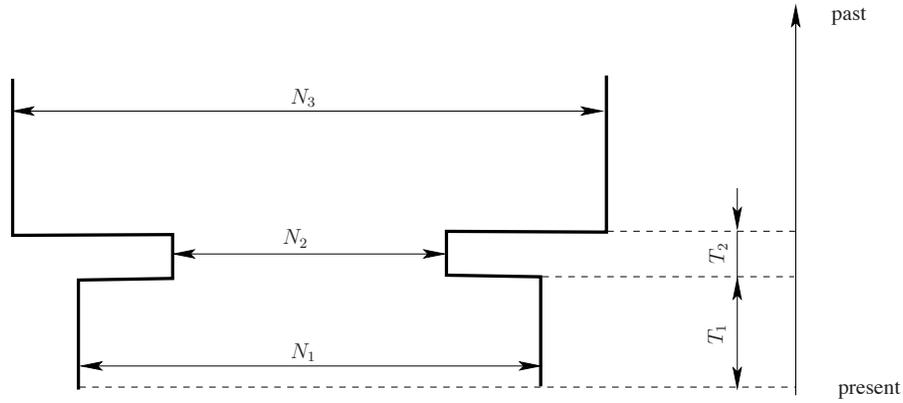}}\hspace*{1cm}
\caption{\label{fig:model} Demographic model. Present population size is $N_1$. In the past, two population-size changes occurred: one at $T_1$ generations ago from $N_1$ to $N_2$ and another one $T_1+T_2$ generations ago from $N_2$ to $N_3$.}
\end{figure}

\section{Materials and methods}\label{sec:methods}

\subsection{Demographic model}\label{sec:model}
We assume a piecewise constant demography with two population-size changes in the past as illustrated in Fig.~\ref{fig:model}. When $N_2<N_1$ and $N_2<N_3$ the demography represents a population bottleneck.  The model of piecewise constant demographies with two population-size changes in the past was considered before \cite{Adams,Marth,Stajich:2005} to capture the main events of the human out-of-Africa expansion \cite{Cavalli,Ramachandran,Liu,TanabeMita:2010,Eriksson:2012}.

In the following we assume a random mating Wright-Fisher diploid population \cite{Wright:1931,Fisher:1930}. We also assume that the population size is large so that the standard coalescent approximation to the Wright-Fisher population can be used \cite{King:1982}.

\subsection{Demography-adjusted tests of neutrality}
\citet{Taj:1989} introduced a test of neutrality which compares two estimators of the scaled mutation rate $\theta=4\mu LN$, with $N$ denoting diploid population size, $\mu$ mutation rate per site, per chromosome, per generation, and $L$ the number of sites in the genomic sequence.
If mutations are neutral, these two estimators have the same expected values. A significant
difference between them indicates a violation of the null assumptions, i.\,e.\, either the population size is varying, or mutations are not neutral (or both). 
Several other tests of neutrality, relying on the same idea and on the same null model, have been proposed since
(\citet{FuLi:1993}, \citet{FayWu:2000}, \citet{Zeng:2006}, \citet{Achaz:2008}). \citet{Achaz:2009} showed that estimators of $\theta$ in any of these tests can be expressed as linear combinations of the SFS, and as instances of a single general formula (see Eq.~(8) in \citet{Achaz:2009}).

We show that this can be further generalised to include demographies with varying population size. Following the notation introduced by \citet{Achaz:2009} and \citet{Ferr:2010}, we write the null site frequency spectrum in the form $\expt{\xi_i}=\xi_i^0\theta$. Here $\xi_i^0=\expt{\xi_i}|_{\theta=1}$ is the expected total branch length of lineages in the gene genealogical tree of the sample that have exactly $i$ leafs. It depends on the sample size $n$ and the parameters of the demography, but not on $\theta$.
It follows that in a sample of size $n$, the SFS provides $n-1$ unbiased estimators $\hat\theta^{(i)}=\xi_i/\xi_i^{0}$. In fact, any linear combination of $\hat\theta^{(i)}$ can be used as an estimator of $\theta$: 
\be\label{eq:theta_j}
\hat{\theta}_\omega=\sum_{i=1}^{n-1}\omega_{i}\hat\theta^{(i)}\,\,,
\ee
where $\omega_i$ are the weights satisfying $\sum_i\omega_i=1$. 
All tests mentioned above compare two different such estimators and are determined only by the difference
$\Omega_i=\omega^{(1)}_i-\omega^{(2)}_i$ of the corresponding weights (listed in Table 1 and 2 of
\citet{Achaz:2009}).

It follows from Eq.~(\ref{eq:theta_j}) that a demography-adjusted test of neutrality, denoted by $T_\Omega$ below, takes the form \cite[their suppl. Eq.~(20)]{Ferr:2010}: 
\be\label{eq:test}
T_\Omega=\frac{\sum_{i=1}^{n-1}\Omega_i\hat\theta^{(i)}}{\sqrt{{\rm Var}\Bigl[\sum_{i=1}^{n-1}
\Omega_i\hat\theta^{(i)}\Bigr]}}\,\,.
\ee 
 The denominator in Eq.~(\ref{eq:test}) for a constant population size is given by \citet[his Eq.~(9)]{Achaz:2009}. For a varying
population size, we calculate analogously (see Appendix~\ref{sec:prob0}): 
\be\label{eq:var0}
{\rm Var}\Bigl[\sum_{i=1}^{n-1} \Omega_i\hat\theta^{(i)}\Bigr]=\theta \sum_{i=1}^{n-1}
\frac{\Omega_i^2}{\xi^0_i}+\theta^2 \sum_{i,j=1}^{n-1}
\frac{\Omega_i}{\xi_i^0} \sigma_{ij}^0 \frac{\Omega_j}{\xi_j^0}\,\,,
\ee
where $\sigma_{ij}^0={\rm Cov}(\xi_i,\xi_j)|_{\theta=1}$ for $i\neq j$, and  $\sigma_{ii}^0=({\rm
Var}(\xi_i)-\expt{\xi_i})|_{\theta=1}$, as defined in \citet{Fu:1995}. Note that, according to its definition,
$\sigma_{ij}^0$ does not depend on $\theta$. In the constant population-size case, it
is a function of sample size $n$ (see \citet{Fu:1995}), and for a non-constant demography it is a function of $n$ and of the parameters of the demography.

As Eq.~(\ref{eq:var0}) shows, an estimate of $\theta$ and of $\theta^2$ is needed to calculate the variance.
\citet{Taj:1989} used the estimator
$\hat\theta_S=\frac{1}{\sum_{k=1}^{n-1}\frac{1}{k}}\sum_{i=1}^{n-1}\xi_i=\frac{1}{\sum_{k=1}^{n-1}\frac{1}{k}}S$
(where $S=\sum_{i=1}^{n-1}\xi_i$ is the number of segregating sites). We extend this definition to an arbitrary null spectrum
by setting $\hat\theta_S=\frac{1}{\sum_{k=1}^{n-1}\xi^0_k}S$. We find that an unbiased estimate of $\theta^2$ based on $\hat\theta_S$ is given by (see
Appendix~\ref{sec:prob0})
\begin{equation}\label{eq:theta2}
\widehat{\theta_S^2}=\frac{\hat\theta_S^2-y_n\hat\theta_S}{1+z_n}\,\,.
\end{equation}
Here, $y_n$ and $z_n$ are given by
\be
y_n=(\sum_{i=1}^{n-1}\xi_i^0)^{-1}~{\rm and}~
z_n=(\sum_{i,j=1}^{n-1}\sigma_{ij}^0)(\sum_{i=1}^{n-1}\xi_i^0)^{-2}\,\,.
\ee
For constant population size $y_n$ and $z_n$ reduce to
\be
y_n=(\sum_{i=1}^{n-1} \frac{1}{i})^{-1}~{\rm and}~z_n=\sum_{i=1}^{n-1} \frac{1}{i^2}(\sum_{i=1}^{n-1}
\frac{1}{i})^{-2}\,\,.
\ee
It is known that estimation of $\theta$ by $\hat\theta_S$ is efficient (i.\,e.\, the estimator has minimal variance) for
small values of $\theta$ \cite{FuLi:1993b}. One can show that this holds for our extended version of $\hat\theta_S$ as
well. In fact, the estimator can become efficient even for high values of $\theta$, if recombination is taken into
account. We note that it is common practise to apply tests, such as Tajima's $D$, to recombining sequences \cite{Akey:2004,Stajich:2005,Carlson} although their derivation neglects recombination.

In our genome scan we encounter rather high values of $\theta$ in the range of $50-100$. In this case
 the first summand in Eq.~(\ref{eq:var0}) can be
neglected. Hence, Eq.~(\ref{eq:test}) can be approximated by \be\label{eq:T_apr}
T_\Omega\approx\frac{\sum_{i=1}^{n-1}
\frac{\Omega_i}{\xi_i^0}\xi_i}{\frac{1}{\sum_{i=1}^{n-1} \xi_i^0}S\,\sqrt{\sum_{i,j=1}^{n-1}
\frac{\Omega_i}{\xi_i^0}\sigma_{ij}^0\frac{\Omega_j}{\xi_j^0}}}
\ee
and the adjustment of the tests to demography with varying population size can be interpreted as a combination of
a modified weighting (via $\xi_i^0$) and scaling (via $\xi_i^0$ and $\sigma_{ij}^0$), yielding an affine linear transformation.

Note, that our adjusted tests co-incide with the original ones if population size is constant. In this case,  expressions for $\xi_i^0$ and $\sigma_{ij}^0$ have been explicitly derived by \citet{Fu:1995}. In
case of varying population size, the corresponding expressions are, in general, unknown. 
For a piecewise constant demography, \citet{Marth} derived an expression for the first moment of the SFS. 
In this study, we use results of \citet{Fu:1995} and of \citet{Eri:10} (see also \citet{Wiehe:2008}) to compute the second moment of the SFS under a piecewise constant demography shown in Fig.~\ref{fig:model}. 
We remark, that this can be done in the same way for the folded SFS (FSFS), i.e. when data cannot be polarized.
The details and the corresponding formulae for the demographic model shown in Fig.~\ref{fig:model}
are given in Appendix~\ref{sec:prob1}.
 
\subsection{Estimating demographic parameters using the SFS}\label{sec:fitting}
We use the analytical expressions for the moments of the SFS under a given demography to compute maximum likelihood (ML) estimates of the parameters of our demographic model. We follow a similar approach as
described in \citet{Adams}, namely we calculate the expected SFS for a large set of plausible parameters and choose the 
parameters with highest likelihood, given the data. If SNPs are assumed to be uncorrelated,
the SFS counts $\xi_1,\ldots,\xi_{n-1}$ are multinomially distributed (conditional on the total number of SNPs $S=\sum_{i=1}^{n-1}\xi_i$), with the parameters given by the expected values of $\xi_i$ \cite{Nielsen:2000,Adams}. 

 Similarly, the probability to observe the FSFS $\eta_1,\ldots,\eta_{\lfloor n/2\rfloor}$ in a sample of $S=\sum_{i=1}^{\lfloor n/2\rfloor}\eta_i$ polymorphic sites is multinomial with
\begin{equation}\label{eq:prob}
{\rm Prob}(\eta_1, \eta_2,\ldots,\eta_{\lfloor n/2\rfloor} |S)=\binom{S}{\eta_1, \eta_2,\ldots,\eta_{\lfloor n/2\rfloor}}\prod_{i=1}^{\lfloor n/2\rfloor}p_i^{\eta_i}\,\,.
\end{equation}
In this case, the parameters $p_i$ are given by:
  \be\label{eq:p_i}
p_i=\frac{\expt{\eta_i}}{\sum_{j=1}^{\lfloor n/2\rfloor}\expt{\eta_j}}\,\,.
\ee
As mentioned in the previous subsection, the expression for $\expt{\xi_i}$ (and thus for $\expt{\eta_i}$) under the
model shown in Fig.~\ref{fig:model} is given in Appendix~\ref{sec:prob1}.

\begin{figure}[t]
\centerline{
\includegraphics{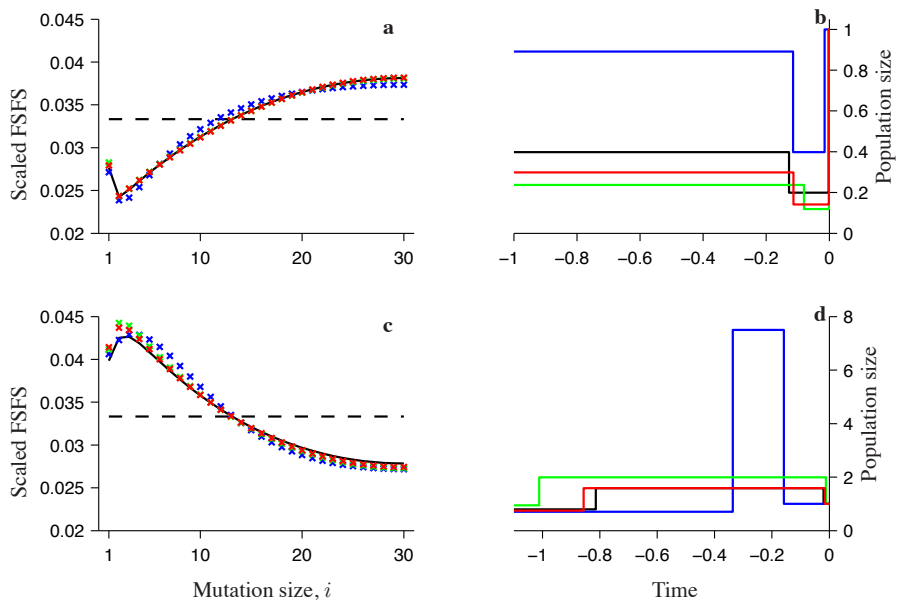}}\hspace*{1cm}
\caption{\label{fig:coal_hist} ({\bf a}), ({\bf c}) Scaled FSFSs  
computed analytically. The spectra are scaled so that, in the constant population-size case, one obtains a constant equal to $1/\lfloor n/2\rfloor$ (shown by dashed lines). Analytical spectra corresponding to the actual underlying demographies (shown by black lines in panels {\bf b} and {\bf d}, respectively) are shown by black lines. The best-fitted spectra estimated using $10^4$ SNPs are shown by blue crosses, green crosses show the best-fitted spectra estimated using $10^5$ SNPs, and red crosses show the best-fitted spectra estimated using $10^6$ SNPs. ({\bf b}) Actual underlying demography (black line) for the spectrum shown in {\bf a} by a black line (recent bottleneck). ({\bf d}) Actual demography (black line) for the spectrum shown in {\bf c} by a black line (past population-size expansion, followed by a recent population-size decline). In {\bf b} and {\bf d} the maximum likelihood histories estimated using $10^4$ SNPs, $10^5$ SNPs, and $10^6$ SNPs are shown by blue, green, and red lines, respectively. The population size is scaled by $N_1$, and the time is scaled by $2N_1$. Sample size used: $n=60$.}
\end{figure}

It is known that different demographies can lead to exactly the same SFS \cite{Myers}. Hence, cases exist in which it is difficult to
distinguish the underlying demographies by their spectra. In order to obtain an estimate for the minimum number of SNPs necessary for reliable inference, we use
coalescent simulations to generate SFSs under two different demographic histories with two population-size changes in the past (see Fig.~\ref{fig:coal_hist}). 
Reconstruction of the ancestral allele via an outgroup is prone to mis-specification, which can substantially bias demography estimation.  We therefore used the folded SFSs (FSFSs) for demography estimation, which is independent of the ancestral allele. 
We simulated $81\cdot 10^6$ independent gene genealogies with $n=60$, and $\theta=0.01$. For such a small value of $\theta$, genealogies rarely contain more than one mutation. For each demography, we determine three resulting FSFSs, one containing $10^4$ SNPs, one with $10^5$ SNPs, and one with $10^6$ SNPs (see circles in Fig.~S1 in Supplementary material). To obtain the FSFSs in a way consistent with practical data sampling, we  randomly select exactly one SNP from randomly chosen genealogies having mutations. Using such spectra, we compute ML-parameters of demographies with two population-size changes in the past. We note that, under the model considered, there are four unknown parameters to be determined. Upon scaling the parameters of the model  ($N_1$, $N_2$, $N_3$, $T_1$, $T_2$) by the present population size $N_1$, the unknown parameters actually are the scaled population sizes $x_i=N_i/N_1$ ($i=2,3$), and the scaled times $t_i$ such that $T_i=\lfloor 2t_iN_1\rfloor$ ($i=1,2$). For the given parameters $x_2$, $x_3$, $t_1$, and $t_2$, the probabilities $p_i$ can be computed using Eqs.~(\ref{eq:xi})-(\ref{eq:xi_last}) in Appendix~\ref{sec:prob1}. Note that the ML-estimation does not depend on the parameter $\theta$, as Eq.~(\ref{eq:p_i}) shows. The ML-demographies are found by  computing ${\rm Prob}(\eta_1,\ldots,\eta_{\lfloor n/2\rfloor} |S_n)$ for a set of candidate parameters: the logarithms of candidate population sizes $x_2$, and $x_3$ are taken from a grid within the  interval $[-2,2]$, and the logarithms of candidate times $t_1$, and $t_2$ are taken from a grid within the  interval $[-3,0]$ (in both cases successive points are equally spaced by $0.025$ units).  Thus, for each population we test in total $121^2\cdot161^2=379,509,361$ combinations of the four unknown demographic parameters. The results are shown in Section~\ref{sec:res}.

We apply this procedure to the FSFSs of ten human populations  (see Table~\ref{tab:1}) to estimate the parameters of the
corresponding piecewise constant demographies with two population-size changes in the past (Fig.~\ref{fig:model}). 
Data were taken from the $1000$ genomes project \cite{1000Genomes}, version $3$ of the release of integrated variant
calls from April $30$th, $2012$.  Variants were filtered by variant type ``SNP''
(i.e. indels excluded). From each population, four (possibly overlapping) subsamples of $30$ individuals were drawn. We used only SNPs from intergenic regions.

\begin{table}
\caption{Populations and the corresponding number of individuals sampled (data from the $1000$ genomes project \cite{1000Genomes}).}
\centering
\begin{tabular}{l|l|c}
     & Population & Sample\\\hline
CEU	 & CEPH individuals & 85 \\
FIN	 & Finnish in Finland & 93\\
GBR	 & British from England and Scotland & 89\\ 
TSI	 & Toscani in Italia & 98\\
CHB	 & Han Chinese in Beijing, China & 97 \\ 
CHS	 & Han Chinese South, China & 100\\
JPT	 & Japanese in Tokyo, Japan & 89\\
ASW	 & African ancestry in Southwest USA & 61\\ 
LWK	 & Luhya in Webuye, Kenya & 97\\
YRI	 & Yoruba in Ibadan, Nigeria & 88\\
\end{tabular}
\label{tab:1}
\end{table}

As explained above, in order to use the analytical formulae for parameter estimation, 
SNPs must be uncorrelated, i.\,e.\, unlinked.
On the other hand, a large amount of SNPs is necessary to render the demography estimation reliable. As a compromise
we collect the SNPs in the following way: from each of the 4 subsamples of $30$ individuals we draw randomly $10^4$ SNPs with the condition
that the minimal physical distance between any pair of SNPs is $5\cdot10^4$ base pairs ($50$ kb). This is repeated $10$
times for each subsample to obtain in total $40$ random spectra. We perform the ML-estimation for each population
by using the average of these $40$ spectra. Results are shown in Section~\ref{sec:res}.

\begin{figure}[t]
\centerline{
\includegraphics{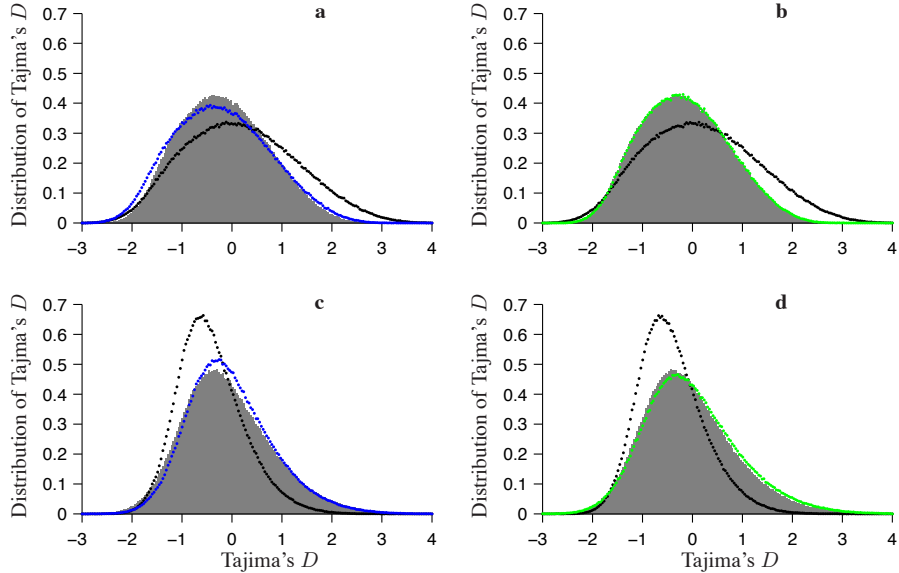}}
\hspace*{1cm}
\caption{\label{fig:coal_Taj} ({\bf a}), ({\bf b}) Numerically computed distributions of Tajima's $D$ for demographic histories shown in Fig.~\ref{fig:coal_hist}{\bf b}. Grey region shows the distribution of Tajima's $D$ adjusted to the actual underlying demography, black circles show the original test and coloured circles show the test adjusted to the maximum likelihood demographies (for a given number of SNPs). Results of  the estimation based on $10^4$ SNPs are shown in panel {\bf a}, and on $10^5$ SNPs in panel {\bf b}. ({\bf c})-({\bf d}) Same as in panels {\bf a}, {\bf b}, respectively, but for demographic histories shown in Fig.~\ref{fig:coal_hist}{\bf d}. Scaled mutation rate used: $\theta=100$. Number of independent gene genealogies simulated: $10^6$.}
\end{figure}

\subsection{Whole-genome scans with demography-adjusted tests of neutrality}
First, we investigate with simulations the error introduced by demography inference. We simulate $10^6$ independent gene
genealogies under two idealized demographies roughly representing the populations CEU and YRI, shown in
Fig.~\ref{fig:coal_hist}{\bf b}, {\bf d} by black lines (recent bottleneck in {\bf b}, and past population-size
expansion followed by a recent decline in {\bf d}). We performed coalescent simulations with
$\theta=100$, corresponding to the values in our genome scan. For each gene genealogy, we compute the distribution
of Tajima's $D$ adjusted to the actual demography, as well as to the estimated demography, and we compare the two.

We perform genome wide computation of Tajima's $D$, Fay $\&$ Wu's $H$ and Zeng's $E$ using the approach by
\citet{Carlson}. We calculate the tests in a sliding window of size $100$ kb and step size 
$10$ kb. Windows containing less than $5$ SNPs
were ignored and we collected about $280,000$ data points. For the tests of Fay $\&$ Wu, and of Zeng it is
necessary to know the ancestral allele. This information
was obtained through a $6$-way alignment of humans and five other primates and is included into the $1000$ genomes data. 
In order to detect putative regions under selection, we distinguished so-called ``contiguous regions of Tajima's
$D$ reduction (CRTR)''. As in \citet{Carlson} we define them as a genomic region of at least 20 consecutive windows, of
which at least 75 $\%$ show a Tajima's $D$ belonging to the $1 \%$ lowest overall values.

\section{Results}\label{sec:res}
\subsection{Test of the maximum likelihood procedure}
In Fig.~\ref{fig:coal_hist}{\bf a}, {\bf c} we show by black lines the analytically computed scaled FSFSs under a recent bottleneck ({\bf a}), that is under a past population-size expansion followed by a recent decline ({\bf c}). 
The spectra are scaled so that in the constant population-size case one obtains a constant value (independent of $i$) equal to $1/\lfloor n/2\rfloor$ (dashed lines in Fig.~\ref{fig:coal_hist}{\bf a}, {\bf c}). 
The demography estimation is based on the spectra obtained using coalescent simulations with $10^4$, or $10^5$, or $10^6$ SNPs (see blue, green, and red circles in Fig.~S1{\bf b}, {\bf d} in Supplementary material). 
By comparing the actual underlying histories to the estimated ones, we find that our ML-procedure works well when using spectra with $\ge 10^5$ SNPs.

In Fig.~\ref{fig:coal_Taj} we show the distributions of Tajima's $D$ adjusted to the ML-demographies shown in Fig.~\ref{fig:coal_hist}{\bf b}, {\bf d} (blue, and green circles). 
For comparison, we also show the distributions of Tajima's $D$ adjusted to the corresponding actual demographies (grey regions), and to the constant population-size history, i.\,e. original Tajima's $D$ (black circles). 
Fig.~\ref{fig:coal_Taj}{\bf a} and {\bf b} show the results based on $10^4$ SNPs, and Fig.~\ref{fig:coal_Taj}{\bf c} and {\bf d} show the results based on $10^5$ SNPs. Our results show that Tajima's $D$ adjusted to the ML-demography coincides well with Tajima's $D$ adjusted to the actual underlying history if the demography estimation is performed using $\ge 10^5$ SNPs (compare Fig.~\ref{fig:coal_Taj}{\bf a} and {\bf c} to Fig.~\ref{fig:coal_Taj}{\bf b} and {\bf d}). 
Note, that while we adjust the tests for the first two moments, demography influences also higher moments. This
leads to a skewness of the adjusted distributions versus the neutral ones as noticed already by
\citet{Wiehe:2008}.
\subsection{Estimated human demographies}

\begin{table}[ht]
\caption{\label{tab:samples} Average and standard deviation (SD) of singletons as an indicator of the
differences between frequency spectra. Compared are four independent drawings of SNPs (each $10^5$ SNPs) out of the same
population subsample with those of different subsamples. A subsample consists of 30 individuals.}\begin{center}
\begin{tabular}{l|cccc}
Population & Intra-sample average & SD & Inter-sample average & SD\\\hline 
CEU & 2029 & 14.0 & 2043 & 25.0\\
FIN & 1894 & 16.9 & 1896 & 18.9\\
GBR & 2062 & 9.4 & 2064 & 17.1\\
TSI & 2165 & 9.5 & 2165 & 9.2\\
CHB & 2039 & 16.2 & 2031 & 23.9\\
CHS & 2048 & 13.3 & 2036 & 52.2\\
JPT & 1955 & 10.8 & 1944 & 16.7\\
ASW & 2837 & 7.3 & 2833 & 23.0\\
LWK & 2665 & 15.3 & 2652 & 71.6\\
YRI & 2352 & 6.6 & 2350 & 24.4\\
\end{tabular}
\end{center}
\end{table}

We now analyze the reliability of the obtained frequency spectra of the human populations. Table \ref{tab:samples}
gives an overview of the variation contained in the empirical FSFSs of the populations. We focus on singletons (mutations of
size $1$) since they represent the most distinctive part of the frequency spectrum between populations. For each
population we compare multiple SNP samplings of the same subsample of $30$ individuals to those of different
subsamples of the same size. It can be seen that our procedure to extract $10^5$ SNPs essentially grasps the
information contained in a specific subsample, since we find only minor changes by repeating it on the same sample. The variation
between different subsamples, which is highest for LWK, may hint at some substructure in a given population. 
The populations CHB, CHS, GBR and CEU are not distinguishable by their amount of singletons (see Table~\ref{tab:samples}), 
but they become distinguishable when doubletons are taken into account (not shown). The difference between CHB and CHS
remains small, though, and their whole frequency spectra are the most similar ones among all populations.
  
Our demography estimation shows (see Fig.~\ref{fig:all} and Table~S1 in
Supplementary material) that the FSFSs of the non-African populations are consistent with a population bottleneck. By
contrast, the FSFS of  the African population ASW is consistent with two population-size
expansions, and the FSFSs of LWK and YRI are consistent with an inverse bottleneck.

\subsection{Neutrality tests adjusted to the estimated human demographies}

\begin{figure}[t]
\includegraphics{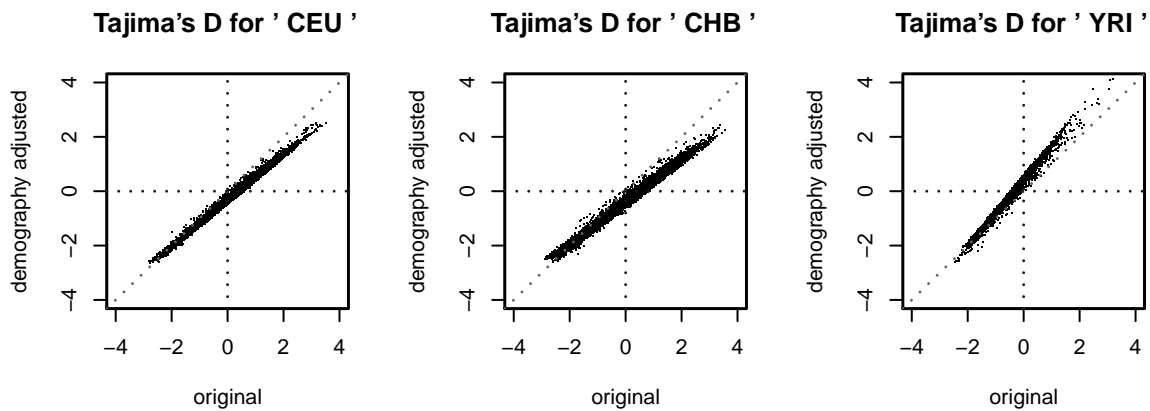}
\caption{\label{fig:scatter}Scatterplots of original vs adjusted tests, for non-overlapping windows;
$\approx 26500$ data points. Fraction of variance explained $R^2\approx0.98$ in all three cases.}
\end{figure}

Figure \ref{fig:scatter} shows  the original test values of Tajima's D plotted against the adjusted ones for nonoverlapping
windows of size $100$kb. The inclusion of demography into the tests basically results 
in an affine linear transformation of the test values (coefficient of determination $R^2\approx 0.99$). Since $\theta$ is large ($\theta>50$ for almost all regions), this observation fits our
theoretical result of Eq.~(\ref{eq:T_apr}). The residuals of a linear regression of the adjusted on the original values
are approximately normally distributed with standard deviation of $\approx 0.07$. This suggests that the scattering observed in the figure should be interpreted as noise and not as a biological phenomenon. Some of the ``outliers'' appear to be due
to windows containing very few SNPs. However, on the other hand, we notice that the residuals of different subsamples
are correlated ($R^2\ge 0.5$) for the same population, but not for different populations. This hints at a possible systematic
effect. The linearity implies that the empirical quantiles of the test statistics are unaffected by the
adjustment.

\begin{figure}[t]
\centerline{
\includegraphics{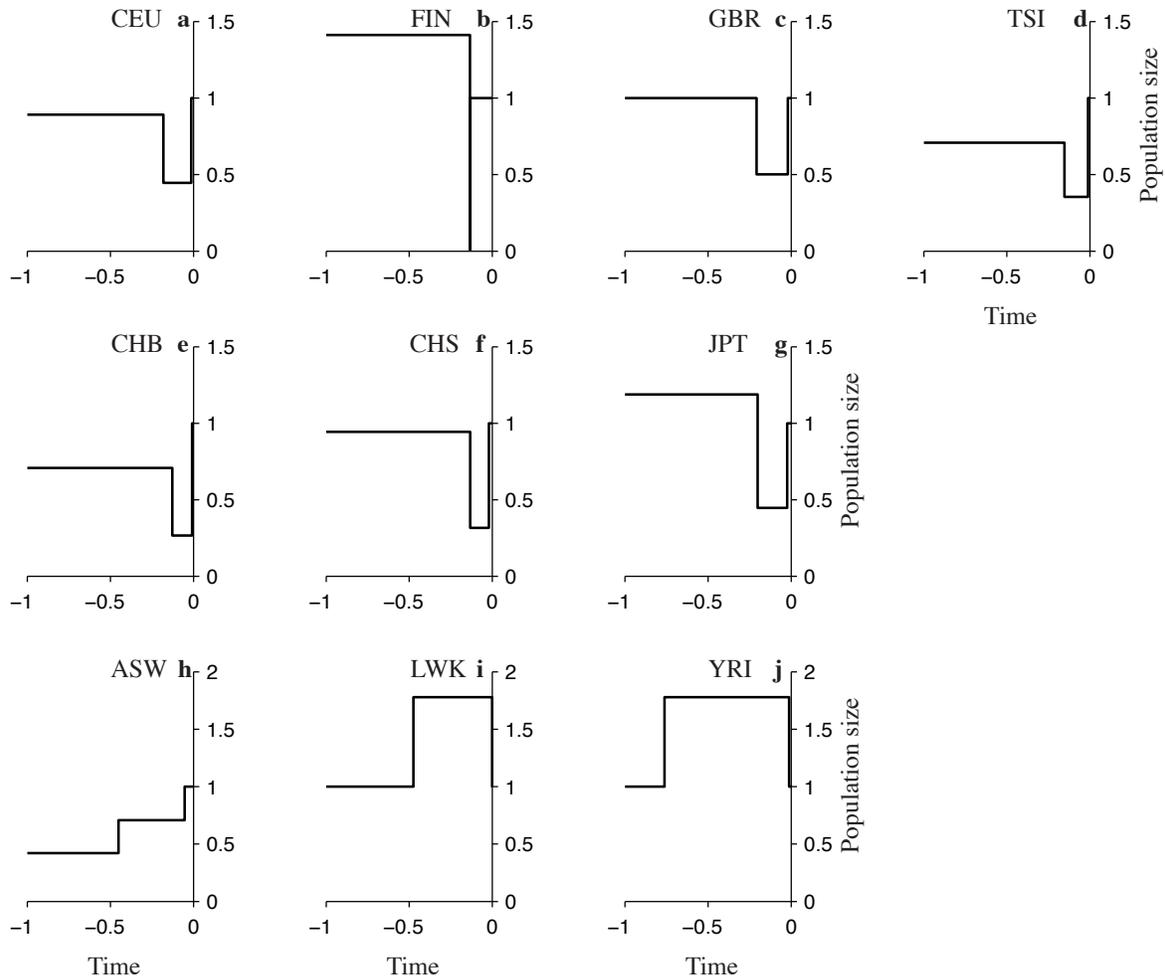}
}\hspace*{1cm}
\caption{\label{fig:all} Estimated demographies for $10$ human populations. Note that the demographies of LWK and YRI have identical shape (inverse bottleneck). However, in both cases the population-size decline is so recent, that it
cannot be seen on this scale. In each panel, the size is scaled by $N_1$, and time is scaled by $2N_1$.
}
\end{figure}

\begin{figure}[t]
\includegraphics[scale=1.1]{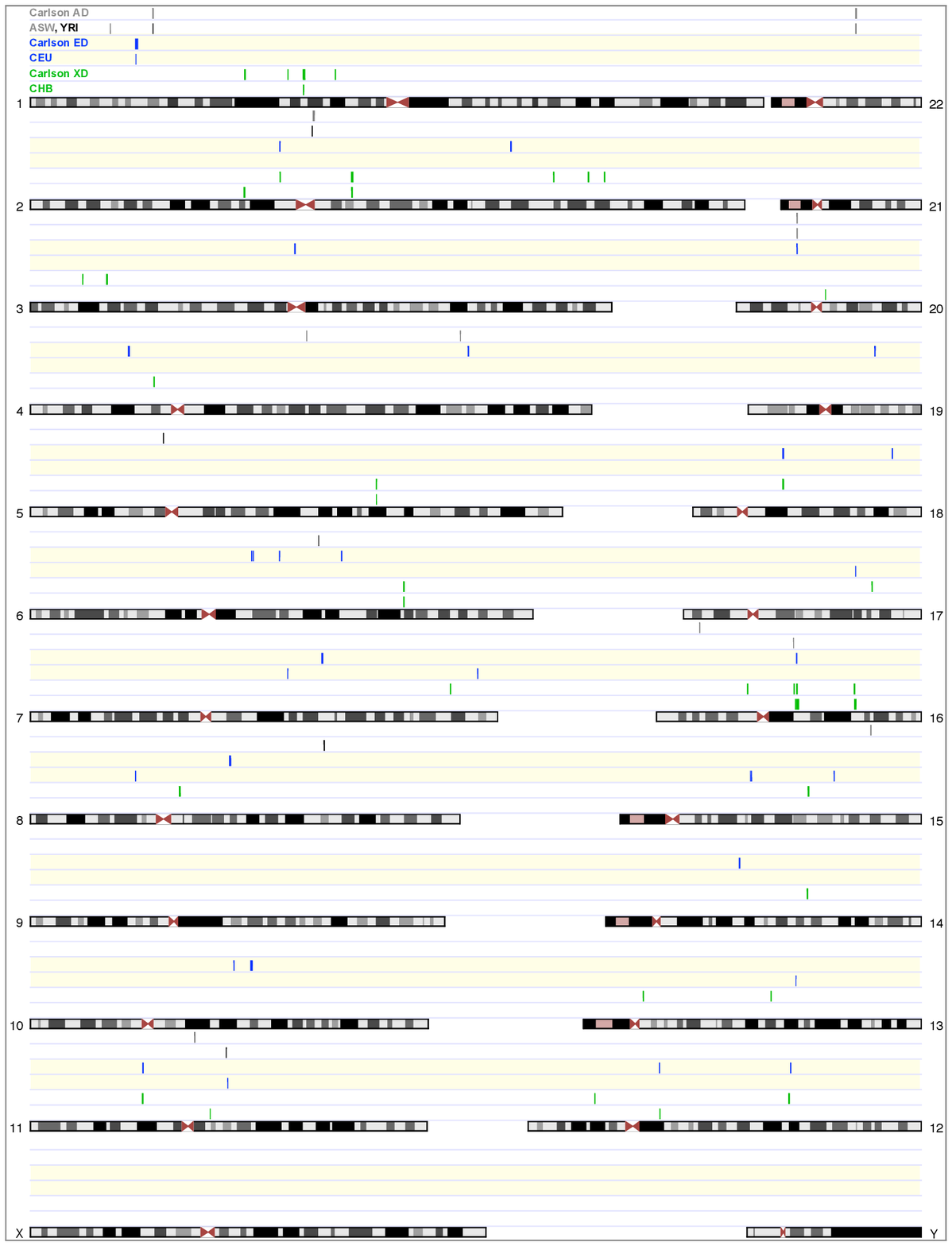}
\caption{\label{fig:CRTR}Contiguous regions of Tajima's D reduction (``CRTR'') from Carlson et al. (2005) 
compared with those derived from our demography-adjusted test. From above to beneath: Carlson: African
descent (gray); ASW (gray) and YRI (black); Carlson: European-descent (blue);
CEU; Carlson: Chinese-descent (green); CHB.
The regions found by Carlson et al. have been translated from hg17 to hg19 coordinates.}
\end{figure}
%

 \subsection{Identifying candidate regions of positive selection}
 We compare Tajima's $D$ between the four subsamples of the same population. The coefficient of determination is
 about $R^2\approx 0.8$ in all populations. The highest correlation between samples from different populations show CHB with CHS
 ($R^2\approx 0.73$), and CEU with GBR ($R^2\approx 0.71$). The lowest correlation show LWK or YRI compared with the
 Asian populations ($R^2\approx 0.1$). We find that CRTRs vary considerably among subsamples of the same population. We
 therefore add a condition and require the test statistic of a particular window to be in the $1\%$-quantile
 simultaneously for all four subsamples. From these windows  we
 construct CRTRs as described above. The additional constraint reduces the number of CRTRs by more than $50\%$. 
 For the populations CEU, CHB and YRI the obtained regions are depicted in
 Figure \ref{fig:CRTR}.  We obtain $7$ ($10$ for adjusted
 test values) CRTRs for population CEU, $10$ ($11$) for CHB and $8$ ($6$) for YRI, respectively. \citet{Carlson},
 using the SNP array data available at that time, obtained $7$ CRTRs for the African, $23$ for the European and $29$ for the
 Chinese population samples which only partially overlap with ours. These differences are caused most likely by the distinct population samples used. In the supplement we list CRTRs of all 10 populations. If the relation between original and adjusted test
 values was linear, their respectively detected regions should be identical. The observed
 differences are probably due to noise which, even if small, leads to split or fused CRTRs.
  
 \section{Discussion and conclusions}\label{sec:conc}
The aim of this study was to incorporate the effects of varying population sizes into SFS-based tests of the neutral evolution hypothesis. We achieved this by adjusting the first two moments of the
site frequency spectrum (SFS) to correspond to a given demography. For populations of constant size the 'adjusted' tests are identical to
the original ones. Our procedure generalises previous results regarding demography-adjustment of Tajima's $D$ \cite{Wiehe:2008}.

When dealing with experimental data, the demography used for adjusting the tests needs to be 
either known from other sources or to be estimated. One method for the estimation is the ML-procedure applied to single nucleotide polymorphisms (SNPs) sampled at physically distant sites, as proposed by \citet{Nielsen:2000} (see also \citet{Adams}). Under this method, individual SNPs are independent from each other and therefore the corresponding SFS counts are multinomially distributed, which simplifies mathematical treatment. Since the parameters of the estimated demography usually differ from those of the real (but generally unknown) demography, we tested by means of  computer simulations how sensitive ML-estimates are with respect to the number of SNPs used for estimation.    
We fitted folded site frequency spectra (FSFSs) simulated under two reference demographies, one being a recent
bottleneck, and the other being a past population-size expansion followed by a recent decline. These two demographies
are instances of a demographic model with two population-size changes in the past. Such a model  is believed to capture the essence \cite{Adams,Marth,Stajich:2005}  of the out-of-Africa expansion of
humans \cite{Cavalli,Ramachandran,Liu,TanabeMita:2010,Eriksson:2012}. Despite its simplicity
four parameters have to be estimated, and therefore a large number of parameter combinations to be tested.
However, it yields exact analytical expressions for the first two moments of the SFS by combining the results of \citet{Fu:1995} with those of \citet{Eri:10}. Note that these expressions are also helpful to find optimal tests of neutrality under piecewise constant demographies \cite{Ferr:2010}. As expected, we found that ML estimation of demography is consistent: the estimated parameters converge to those of the true demography with increasing number of SNPs. The spectrum corresponding to the ML-demography is almost indistinguishable from the spectrum corresponding to the real underlying demography if the estimation is based on more than $100,000$ SNPs. We confirmed this finding for our two reference demographies by comparing Tajima's $D$ adjusted to the actual underlying demography, with that adjusted to the ML-demography.

After confirming the validity of the ML-procedure, we applied our method to disentangle the effects of selection and
demography using data from the $1000$ genomes project \cite{1000Genomes}. We sampled the FSFSs of ten human populations from physically distant  intergenic regions  (presumably neutral \cite{Adams}) in order to estimate the  ML-parameters of the piecewise constant demographic model with two population-size changes in the past
allowing for population size parameter changes of at most two orders of magnitude \cite{Marth}. The time parameters were allowed to vary by three orders of magnitude (i.e. from $-3$ to $0$ on logarithmic scale). The lower boundary for the times corresponds to only $10$ generations (that is, $200-250$ years, under the assumption that a human generation time is $20-25$ years \cite{Marth}). This is a very short time, and we do not expect that demographic changes occurring on even shorter timescales would be detected by the site frequency spectra (since the process of  mutations is slow). In fact,   Eq.~(\ref{eq:xi1}) in Appendix~\ref{sec:prob1} shows that in the limit $t_1\ra 0$, the SFS, and therefore the FSFS, corresponds to that of a  two-stage demography with population  size equal to $x_2$ in the first stage, and $x_3$ in the second stage. The upper boundary for the times was chosen to coincide with the emergence of anatomically modern humans about 200,000 years ago (see \citet{Cavalli} and references therein).

Our results are mainly consistent with the results of \citet{Adams} and on \citet{Marth}: the ML-demographies of non-African populations correspond to a bottleneck, and the ML-demography of one of the sampled African populations (ASW) corresponds to two subsequent population-size expansions. The FSFSs of the remaining two African populations (LWK and YRI) gave rise to demographies with a distant population-size expansion followed by a population-size decline. 

In order to detect regions under selection, we computed genome-wide values of three tests of neutrality, by scanning
over sliding windows with $100$ kb, as proposed by \citet{Carlson}. We find that the distributions of the adjusted
tests are very similar to each other, suggesting that the differences between the original distributions can
be explained to a large part by demography. We find that the adjusted test values are essentially affine linear
transformations of the original ones. This leads to largely identical quantiles and, consequently, identical candidate
regions for selection. Our results show that it is valid to use the original tests in order to detect
selection as long as the empirical distribution of test values of the whole genome is used as reference. The adjusted
values are however useful, as they facilitate direct comparisons of test values from different populations.
Therefore we provide our genome scans of both original and adjusted tests as tracks for the UCSC genome browser.

\citet{Carlson} calculated the correlation between Tajima's D derived from SNP array data with that from
resequenced genes from the same individuals. We compare the former with our values for all
windows and find a lower correlation,  most likely due to distinct
population samples. As a consequence, also the candidate regions of selection
show only modest  overlap. We find that the specification of these
regions as long consecutive stretches of extremely low Tajima's $D$, while in general useful, is sensitive to minor changes in single windows. We therefore try to make this concept more robust by requiring windows to belong to the respective lower $1\%$-quantile in several subsamples of the same population. This reduces drastically the amount of candidate regions. The differences between regions identified using original vs adjusted values is the result of the slight scattering of the transformation which splits some contiguous
regions and fuses others.

Concerning the validity and consistency of our results, our main point is
that the inference of demography by the ML-approach is very sensitive to minor changes in the
frequency spectrum. \citet{Myers} even stated, that the (theoretical) existence of very different
demographies with exactly the same frequency spectrum precludes such an inference altogether.  Our results do not support this overly pessimistic view. Rather, we find that ML-parameter estimation of an, admittedly, simple
demographic model is consistent.

We emphasize, that the adjustment of the tests relies on the absolute values of the inferred moments ($\xi_i^0$ and
$\sigma^0_{ij}$) which are a function of the entire demography not just of quantities (e.g. $\theta$) at present time.
In particular, we observe, that different demographies with similar frequency spectrum can in principle lead to
different variances of the adjusted tests. 

As is common practise, we ignored recombination, although it is known that recombination reduces the variance of the
tests considered. Since recombination is not uniform accross the genome, neglecting it causes a distortion of the test
distributions. However, the demography-adjusted tests studied here serve as a basis for further work 
in which recombination and
rate inhomogeneity across genomes is taken into account. 

The program used to calculate the adjusted test statistics is
available as C++ source code on \url{http://ntx.sourceforge.net/} and
tracks for the UCSC browser containing test values (original as well as adjusted) for all ten populations are available at \url{http://jakob.genetik.uni-koeln.de/data/}\,.

\bigskip

\textit{Acknowledgements.} This work was financially supported by grants from Vetenskapsr\aa det, from the G\"oran Gustafsson Foundation for Research in Natural Sciences and Medicine, through the platform ``Centre for Theoretical Biology" and from CeMEB at the University of Gothenburg to BM, and by a grant of the German Science Foundation (DFG-SFB680) to TW.  

\newpage
\bibliographystyle{elsarticle-harv}
\bibliography{population_gen}

\newpage
\section*{Appendix A: The denominator of demography-adjusted tests of neutrality based on the SFS}\label{sec:prob0}
As explained in Section~\ref{sec:methods}, all tests of neutrality based on the SFS can be expressed using a general form, Eq.~(\ref{eq:test}). The numerator of Eq.~(\ref{eq:test}) depends on the first moment of the SFS under a given demography. Similarly, the denominator of Eq.~(\ref{eq:test}) depends on the second moment of the SFS under a given demography. 
We find:  
\begin{eqnarray}
{\rm Var}\Bigl[\sum_{i=1}^{n-1}
\Omega_i\hat\theta^{(i)}\Bigr]
&=&{\rm Var}\Bigl[\sum_{i=1}^{n-1}\Omega_i
\frac{\xi_i}{\xi_i^0}\Bigr]\nonumber\\
&=&\sum_{i=1}^{n-1} {\rm Var}(\Omega_i
\frac{\xi_i}{\xi_i^0})+\underset{i\neq j}{\sum_{i,j=1}^{n-1}}{\rm Cov}(\Omega_i
\frac{\xi_i}{\xi_i^0},\Omega_j\frac{\xi_j}{\xi_j^0})\nonumber\\
&=&\sum_{i=1}^{n-1} (\frac{\Omega_i}{\xi_i^0})^2 {\rm Var}(\xi_i)+\underset{i\neq j}{\sum_{i,j=1}^{n-1}}\frac{\Omega_i}{\xi_i^0}{\rm Cov}(\xi_i,\xi_i)\frac{\Omega_j}{\xi_j^0}\nonumber\\
&=&\theta\sum_{i=1}^{n-1}\Omega_i^2\frac{1}{\xi_i^0}+ \theta^2
\sum_{i=1}^{n-1} (\frac{\Omega_i}{\xi_i^0})^2 \sigma_{ii}^0+ \theta^2
\underset{i\neq j}{\sum_{i,j=1}^{n-1}}\frac{\Omega_i}{\xi_i^0} \sigma_{ij}^0\frac{\Omega_j}{\xi_j^0}\nonumber\\
&=&\theta \sum_{i=1}^{n-1} \Omega_i^2
\frac{1}{\xi_i^0}+\theta^2\sum_{i,j=1}^{n-1}\frac{\Omega_i}{\xi_i^0}\sigma_{ij}^0\frac{\Omega_j}{\xi_j^0}\nonumber\\\,\,.
\label{eq:var}
\end{eqnarray}
Here one has $\sigma^0_{ij}={\rm Cov}(\xi_i,\xi_j)|_{\theta=1}$, for $i\neq j$, and  $\sigma^0_{ii}=({\rm
Var}(\xi_i)-\expt{\xi_i})|_{\theta=1}$. Eq.~(\ref{eq:var}) corresponds to Eq.~(\ref{eq:var0}) given in the main text. 
Note that for the constant population size one has $\xi_i^0=1/i$, and $\sigma^0_{ij}$ is given by \citet{Fu:1995}. Thus,
Eq.~(\ref{eq:var}) reduces to Eq.~(9) in \citet{Achaz:2009}.

In order to compute Eq.~(\ref{eq:var}) using the observed spectrum, one needs to have an estimate of $\theta^2$. 
For a given estimator of $\theta$, that is based on weights
$\omega_1,\ldots,\omega_{n-1}$, i.e. $\hat\theta_\omega=\sum_{i=1}^{n-1}\omega_i\xi_i/\xi_i^0$, it holds
$$\expt{\hat\theta_\omega^2}=\text{Var}[\hat\theta_\omega]+\expt{\hat\theta_\omega}^2=\theta\;
\sum_{i=1}^{n-1} \frac{\omega_i^2}{\xi^0_i}+\theta^2\; \sum_{i,j=1}^{n-1}
\frac{\omega_i}{\xi_i^0}\sigma_{ij}^0\frac{\omega_j}{\xi_j^0}+\theta^2=y_n 
\theta + (1 + z_n)\theta^2,$$ 
with
\be
y_n=\sum_{i=1}^{n-1}\frac{\omega_i^2}{\xi_i^0}~{\rm and}~z_n=\sum_{i,j=1}^{n-1}
\frac{\omega_i}{\xi_i^0}\sigma_{ij}^0\frac{\omega_j}{\xi_j^0}\,\,.
\ee
It follows that $$\expt{\hat\theta_\omega^2}-y_n \expt{\hat\theta_\omega}=\theta^2(1+z_n).$$ 
Solving the latter with respect to $\theta^2$ yields:
$$\theta^2=\frac{\expt{\hat\theta_\omega^2}-y_n
\expt{\hat\theta_\omega}}{1+z_n}.$$
Hence, as an estimator for $\theta^2$ we take 
\be
\widehat{\theta_\omega^2}=\frac{\hat\theta_\omega^2-y_n
\hat\theta_\omega}{1+z_n}\,\,.
\ee
This expression corresponds to Eq.~(\ref{eq:theta2}) given in the main text.

\section*{Appendix B: The first two moments of the SFS}\label{sec:prob1} 
In this appendix we compute the first two moments of the SFS, $\expt{\xi_i}$ and $\expt{\xi_i\xi_j}$, under a varying population size. We consider a large, well mixed, randomly mating diploid Wright-Fisher population with a varying population size. We assume that mutations accumulate according to the infinite sites model at rate $\mu$ per generation per site. The scaled mutation rate, $\theta$, per genetic sequence of length $L$ is given by $\theta=4\mu N_1 L$, where $N_1$ denotes the present population size. We consider the SFS corresponding to gene genealogies of $n$ individuals. The scaled time during which gene genealogies have exactly $k\le n$ lines is denoted by $\tau_k$ below (i.\,e.\, $\tau_k$ stands for $\lfloor 2 N_1 \tau_k\rfloor$ generations).

The first two moments of the SFS can be expressed as \cite{Fu:1995}
\begin{align}\label{eq:xi1}
\expt{\xi_i}&=\frac{\theta}{2}\sum_{k=2}^{n}k p(k,i)\expt{\tau_k}\,\,,\\
\expt{\xi_i\xi_j}&=\delta_{i,j}\sum_{k=2}^n kp(k,i)\left(\frac{\theta}{2}\expt{\tau_k}+\frac{\theta^2}{4}\expt{\tau_k^2}\right)\nonumber\\
\qquad\qquad&+\frac{\theta^2}{4}\left\{\sum_{k=2}^n k(k-1)p(k,i;k,j)\expt{\tau_k^2}+\sum_{k<m}^n k m  \left(p(k,i;m,j)+p(k,j;m,i)\right) \expt{\tau_k\tau_m}\right\}\,\,.\label{eq:xi2}
\end{align}
where
\begin{align}
\delta_{i,j}&=\begin{cases}
1,&{\rm for}~i=j\,\,,\\
0,&{\rm for}~ i\neq j\,\,,
\end{cases}\\
p(k,i)&=\frac{\binom{n-k}{i-1}}{\binom{n-1}{i}}\frac{k-1}{i}\,\,,\\
p(k,i;k,j)&=
\begin{cases}
\frac{\binom{n-i-j-1}{k-3}}{\binom{n-1}{k-1}},&{\rm for}~k>2\,\,,\\
p(k,i),&{\rm for}~{k=2},~{\rm and}~i+j=n\,\,,\\
0,&{\rm for}~{k=2},~{\rm and}~i+j\neq n\,\,,
\end{cases}\\
p(k,i;m,j)&=\left(\delta_{\lfloor i/j\rfloor,0}+\delta_{i,j}\right)p_a(k,i;m,j)+\left(\delta_{\lfloor(i+j)/n\rfloor,0}+\delta_{i+j,n}\right)p_b(k,i;m,j)\,\,.\label{eq:pa_b}
\end{align} 
The probabilities $p_a(k,i;m,j)$, and $p_b(k,i;m,j)$ in Eq.~(\ref{eq:pa_b}) are \cite{Fu:1995}
\begin{align}
p_a(k,i;m,j)&=
\begin{cases}
\sum_{t=2}^{{\rm min}(m-k+1,i-j+1)} \frac{\binom{m-k}{t-1}}{\binom{m-1}{t}}\frac{k-1}{m}\frac{\binom{i-j-1}{t-2}\binom{n-i-1}{m-t-1}}{\binom{n-1}{m-1}},&{\rm for}~j<i\,\,\\
\frac{k-1}{m(m-1)}\frac{\binom{n-i-1}{m-2}}{\binom{n-1}{m-1}},&{\rm for}~i=j\,\,,
\end{cases}\\
p_b(k,i;m,j)&=\begin{cases}
\sum_{t=1}^{{\rm min}(m-2,m-k+1,i)}\frac{\binom{m-k}{t-1}}{\binom{m-1}{t}}\frac{(k-1)(m-t)}{tm}\frac{\binom{i-1}{t-1}\binom{n-i-j-1}{m-t-2}}{\binom{n-1}{m-1}},&{\rm for}~k>2\,\,\\
\frac{1}{jm}\frac{\binom{n-m}{j-1}}{\binom{n-1}{j}},&{\rm for}~k=2,~{\rm and}~i+j=n\,\,.\label{eq:xi_end}
\end{cases}
\end{align}
In the limit $\theta\ra 0$, Eq.~(\ref{eq:xi2}) reduces to:
\be
\expt{\xi_i^2}=\frac{\theta}{2}\expt{\xi_i},~{\rm and}~\expt{\xi_i\xi_{j\neq i}}=0~{\rm for}~\theta\ra0\,\,.
\ee
In other words, in this limit the SFS counts are  multinomially distributed, as explained in Section~\ref{sec:methods}.

For constant population size, it follows from Eq.~(\ref{eq:xi1}) that $i\expt{\xi_i}=\theta$, independently of $i$. By contrast, for demographic history shown in Fig.~\ref{fig:model},  this is not true. Using the results of \citet{Eri:10}, in this case we find:
\be\label{eq:xi}
\expt{\xi_i}=\frac{\theta}{2}\sum_{m_1=2}^{n} a^{(ni)}_{m_1} f_{m_1}\,\,,~{\rm for}~i=1,\ldots,n-1\,\,,
\ee
where $a^{(ni)}_{m_1}$, and $f_{m_1}$ are:
\begin{align}
a^{(ni)}_{m_1}&=\sum_{k=2}^{m_1} k c_{nkm_1} p(k,i)\,\,,\\
f_{m_1}&= b_{m_1}^{-1}\left(1-(1-x_2){\rm e}^{-b_{m_1}\, t_1}+\left(x_3-x_2\right){\rm e}^{-b_{m_1}t_1}{\rm e}^{-b_{m_1}\,s_2}\right)\,\,.\label{eq:xi_last}
\end{align} 
Here, $x_2=N_2/N_1$, $x_3=N_3/N_1$, $s_2=t_2/x_2$, $b_{m_1}=\binom{m_1}{2}$, and $c_{nkm_1}$ is given by Eq.~(11) in \citet{Eri:10}. 
This result is consistent with Eq.~(1) in \citet{Marth}, assuming $M=3$ in the model of \citet{Marth}.

In what follows, we list our results for $\expt{\xi_i \xi_j}$ under the demographic history shown in Fig.~\ref{fig:model}.  
We find:
\be\label{eq:xi_ij}
\expt{\xi_i\xi_j}=\delta_{i,j}\left(\expt{\xi_i}+\frac{\theta^2}{4}\sum_{m_1=2}^n \sum_{k=2}^{m_1}a^{(nij)}_{m_1 k} f_{m_1k}\right)+
\frac{\theta^2}{4}\sum_{m_1=2}^n\left(\sum_{k=2}^{m_1} g^{(nij)}_{m_1 k}f_{m_1 k}+\sum_{m_2=2}^{m_1} h^{(nij)}_{m_1 m_2} f_{m_1 m_2}\right)\,\,,
\ee
where
\begin{align}
a^{(nij)}_{m_1 k}&=2 k c_{nk m_1} c_{kkk} p(k,i)\,\,,\\
g^{(nij)}_{m_1 k}&=2 k (k-1) c_{nk m_1} c_{kkk}p(k,i;k,j)\,\,\\
h^{(nij)}_{m_1 m_2}&=\sum_{l=m_2}^{m_1} l c_{nlm_1}\sum_{k=2}^{m_2} k c_{l k m_2} [p(k,i;l,j)+p(k,j;l,i)]\,\,.
\end{align}
For the terms $f_{m_1,m_2}$ in Eq.~(\ref{eq:xi_ij}), we consider separately the cases $m_1\neq m_2$, and $m_1=m_2$. For the case $m_1\neq m_2$, we find
\begin{eqnarray}
f_{m_1 m_2} &=& \nonumber\\
& & \frac{1}{ b_{m_2}}\Bigg\{\frac{1-{\rm e}^{-b_{m_1}t_1}\big[1-x_2^2+\left(x_2^2-{x_3}^2\right){\rm e}^{-b_{m_1}s_2}\big]}{b_{m_1}}
\nonumber \\
& & -\big[1-x_2+(x_2-x_3){\rm e}^{-b_{m_2}s_2}\big]\frac{{\rm e}^{-b_{m_2}t_1}-{\rm e}^{-b_{m_1}t_1}}{b_{m_1}-b_{m_2}}
\nonumber \\ & &
+x_2(x_3-x_2) {\rm e}^{-b_{m_1}t_1}\frac{{\rm e}^{-b_{m_2}s_2}-{\rm e}^{-b_{m_1}s_2}}{b_{m_1}-b_{m_2}}
\Bigg\}\,\,.
\end{eqnarray}
For the case $m_1=m_2$, we obtain:
\begin{eqnarray}
f_{m_1 m_1}&=& \nonumber \\
& & \frac{1}{ b_{m_1}}\Bigg\{\frac{1-{\rm e}^{-b_{m_1}t_1}\big[1-x_2^2+\left(x_2^2-{x_3}^2\right){\rm e}^{-b_{m_1}s_2}\big]}{b_{m_1}} \nonumber \\ 
& &
-t\big[1-x_2+(x_2-x_3){\rm e}^{-b_{m_2}s_2}\big]{\rm e}^{-b_{m_1}t_1} \nonumber \\ 
& &
+x_2(x_3-x_2) s_2{\rm e}^{-b_{m_1}t_1}{\rm e}^{-b_{m_1}s_2}
\Bigg\}\,\,.  \label{eq:last}
\end{eqnarray}
Eqs.~(\ref{eq:xi})-(\ref{eq:xi_last}) are used to find the demographic parameters that correspond to empirical data in terms of the maximum likelihood approach. Eqs.~(\ref{eq:xi_ij})-(\ref{eq:last}) are used to compute the tests of neutrality under the demographies found. The results are shown in Section~\ref{sec:res}.

~
\protect\newpage
~

\input{supplement}

\end{document}

%% file: supplement.tex
\pagebreak

\newcommand{\beginsupplement}{%
        \setcounter{table}{0}
        \renewcommand{\thetable}{S\arabic{table}}%
        \setcounter{figure}{0}
        \renewcommand{\thefigure}{S\arabic{figure}}%
     }


\protect \newpage
\beginsupplement
\section*{Supplementary Material}

\bigskip

\bigskip

\begin{figure}[th]
\centerline{
\includegraphics{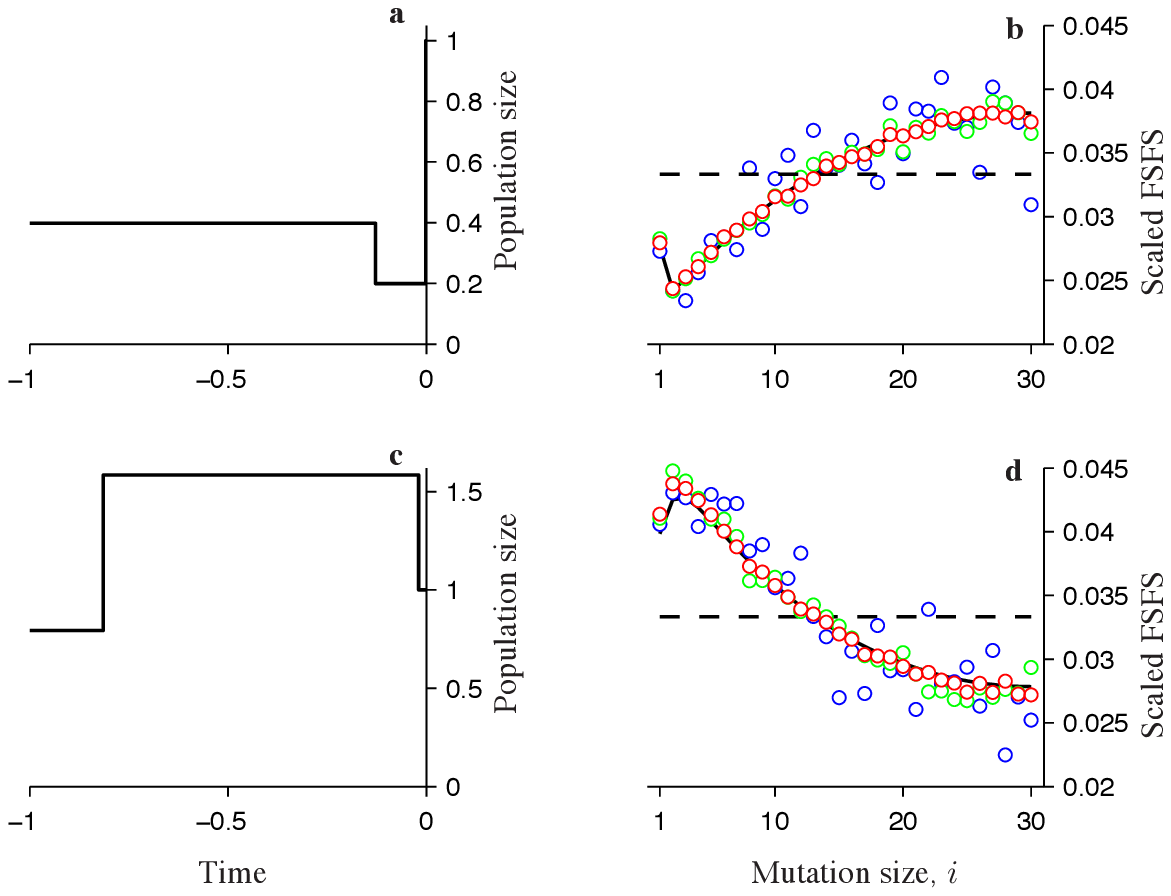}}
\hspace*{1cm}
\caption{\label{fig:coal_spect} ({\bf a}), ({\bf c}) Reference demographic histories (recent bottleneck in {\bf a}, and a past population-size expansion followed by a recent decline in {\bf c}).  ({\bf b}), ({\bf d}) Scaled FSFSs 
computed analytically (black lines), together with the spectra obtained using our coalescent simulations containing $10^4$ SNPs (blue circles), $10^5$ SNPs (green circles) and $10^6$ SNPs (red circles). Each spectrum is obtained by sampling one SNP from gene genealogies that have at least one mutation. The spectra are scaled so that in the constant population-size case, one obtains a constant equal to $1/\lfloor n/2\rfloor$ (see dashed lines). Sample size: $n=60$. Scaled mutation rate used: $\theta=0.01$. Number of independent gene genealogies simulated: $2\cdot10^6$.} 
\end{figure}

\begin{figure}[th]
\centerline{
\includegraphics{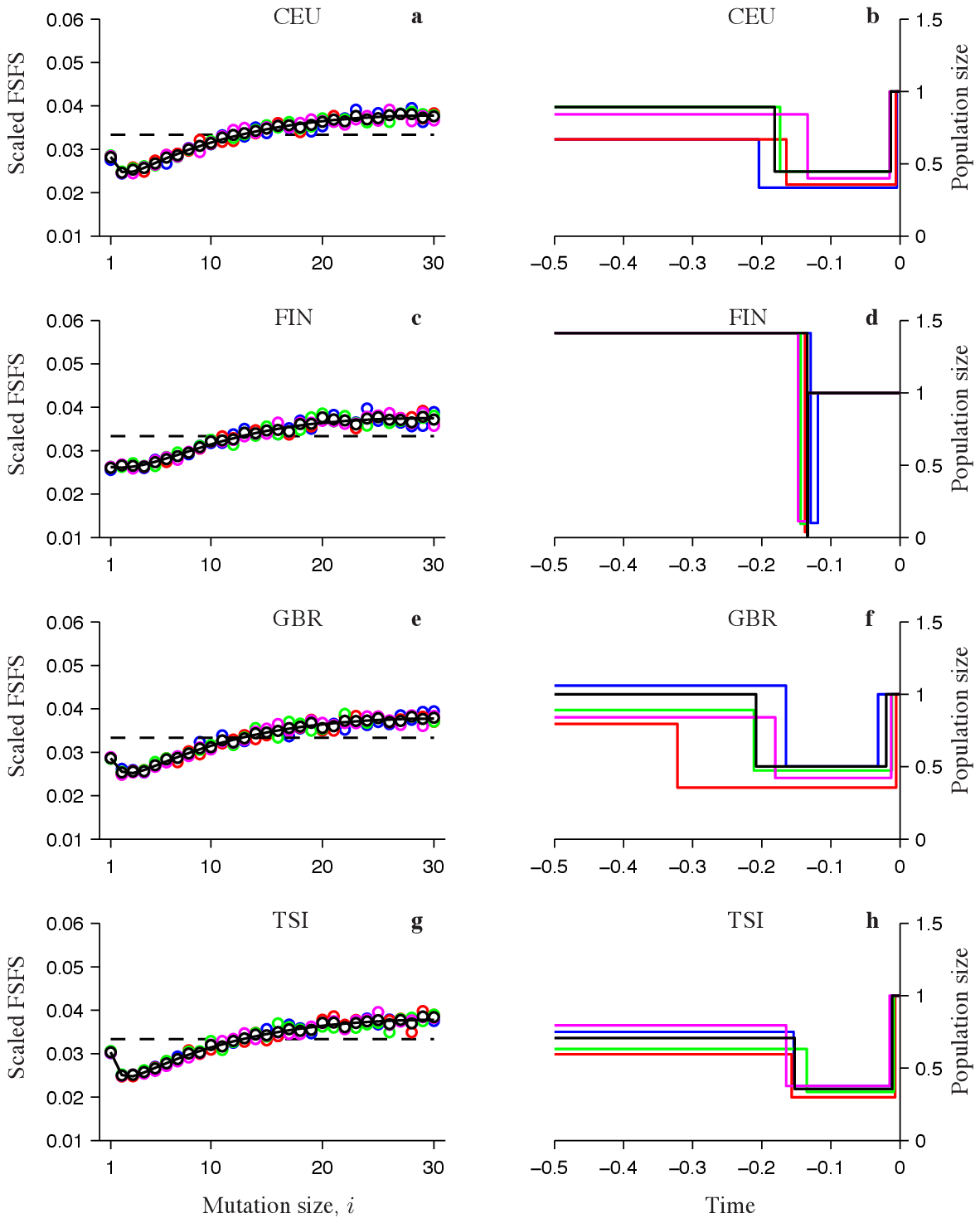}
}\hspace*{1cm}
\caption{\label{fig:eur2} ({\bf a}), ({\bf c}), ({\bf e}), ({\bf g}) Blue, red, green, and magenta circles show four empirically obtained scaled FSFSs 
for the four sampled European populations CEU ({\bf a}), FIN ({\bf c}), GBR ({\bf e}), and TSI ({\bf g}). The spectra are scaled so that in the constant population-size case one obtains a constant equal to $1/\lfloor n/2\rfloor$ (shown by dashed lines). For each population black circles correspond to the spectrum obtained upon averaging over the forty sampled spectra. The corresponding best-fitted scaled spectra are shown by black lines. ({\bf b}), ({\bf d}), ({\bf f}), ({\bf h}) Best-fitted histories corresponding to the empirical spectra (demographies are coloured to match their fitted spectra).}
\end{figure}

\begin{figure}[th]
\centerline{
\includegraphics{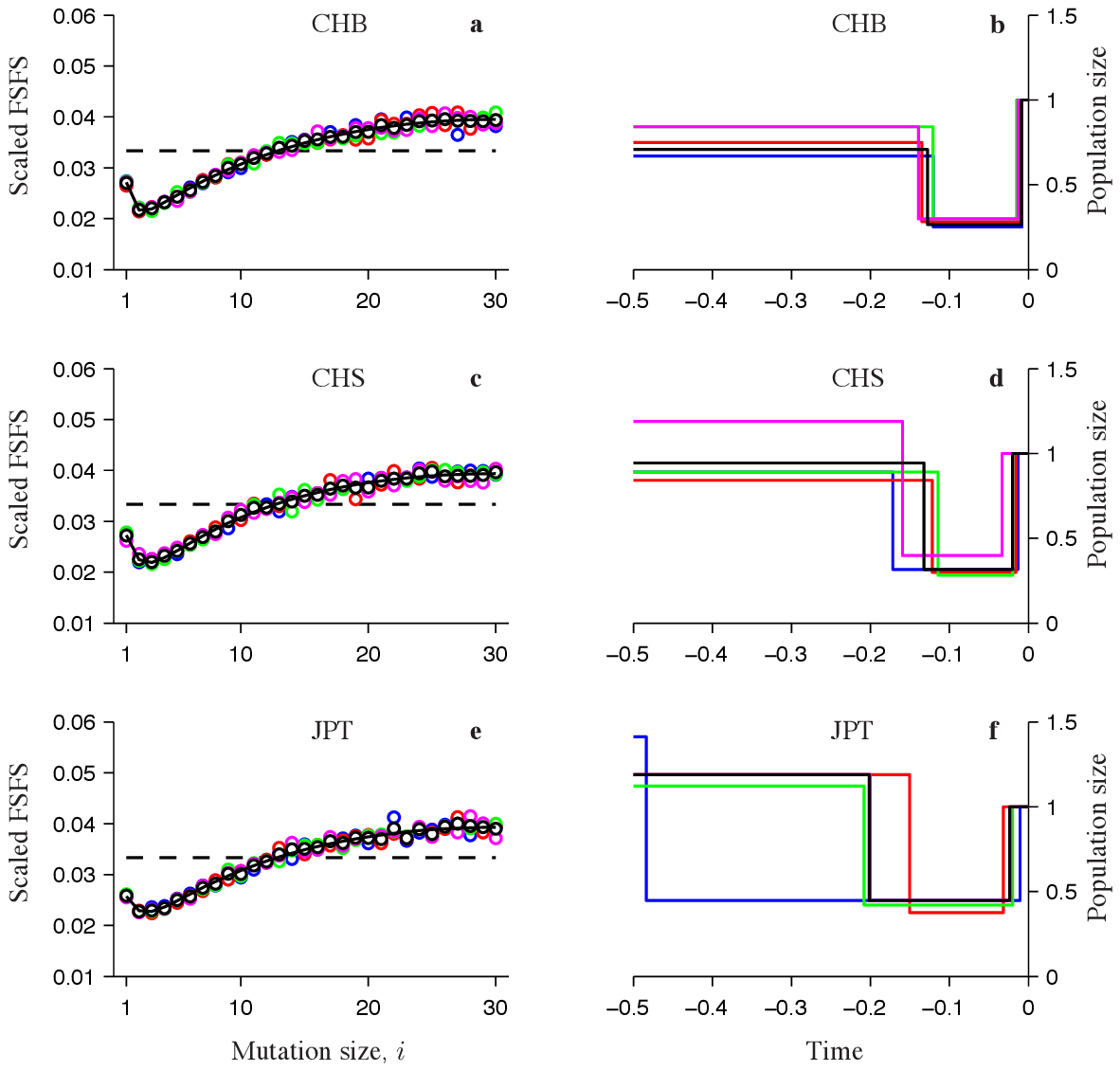}
}\hspace*{1cm}
\caption{\label{fig:asi2} Same as in Fig.~\ref{fig:eur2} but for the populations with Asian ancestry.}
\end{figure}

\begin{figure}[th]
\centerline{
\includegraphics{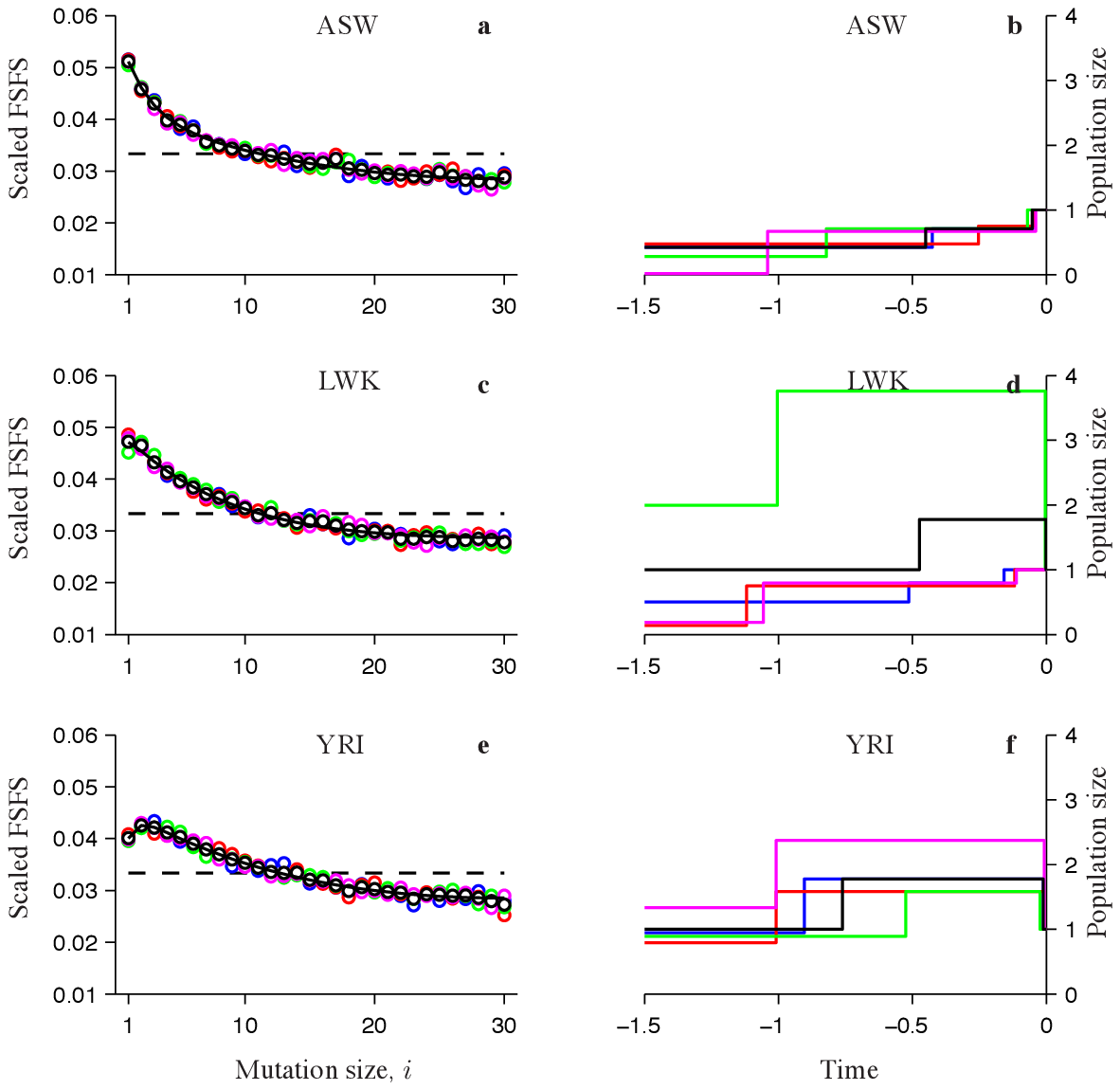}
}
\hspace*{1cm}
\caption{\label{fig:afr2} Same as in Fig.~\ref{fig:eur2} but for the populations with African ancestry. }
\end{figure}

\begin{table}[th]
\caption{Estimated demographic parameters using empirical spectra (the spectra are shown as black circles in Figs.~\ref{fig:eur2}-\ref{fig:afr2}).}
\centering
\begin{tabular}{l|r|r|r|r}
Population&$\log(t_1)$&$\log(t_2)$&$\log(x_2)$&$\log(x_3)$\\
\hline
\hline
CEU &$-1.875$&$-0.775$&$-0.35$&$-0.05$\\
FIN &$-0.875$&$-2.975$&$-2$&$0.15$\\
GBR &$-1.7$&$-0.725$&$-0.3$&$0$\\
TSI &$-1.95$&$-0.85$&$-0.45$&$-0.15$\\
\hline
CHB &$-2.05$&$-0.925$&$-0.575$&$-0.15$\\
CHS &$-1.7$&$-0.95$&$-0.5$&$-0.025$\\
JPT &$-1.625$&$-0.75$&$-0.35$&$0.075$\\
\hline
ASW &$-1.275$&$-0.4$&$-0.15$&$-0.375$\\
LWK &$-3$&$-0.325$&$0.25$&$0$\\
YRI &$-1.925$&$-0.125$&$0.25$&$0$\\
\end{tabular}
\label{tab:2}
\end{table}

\newpage
\begin{figure}[th]
\includegraphics[scale=0.8]{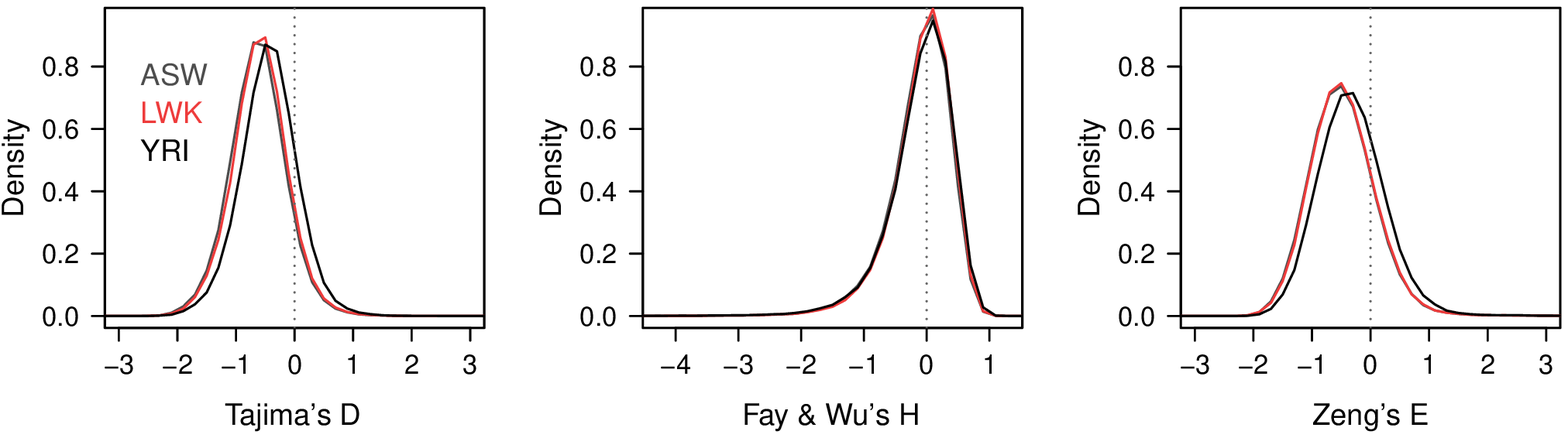}
\includegraphics[scale=0.8]{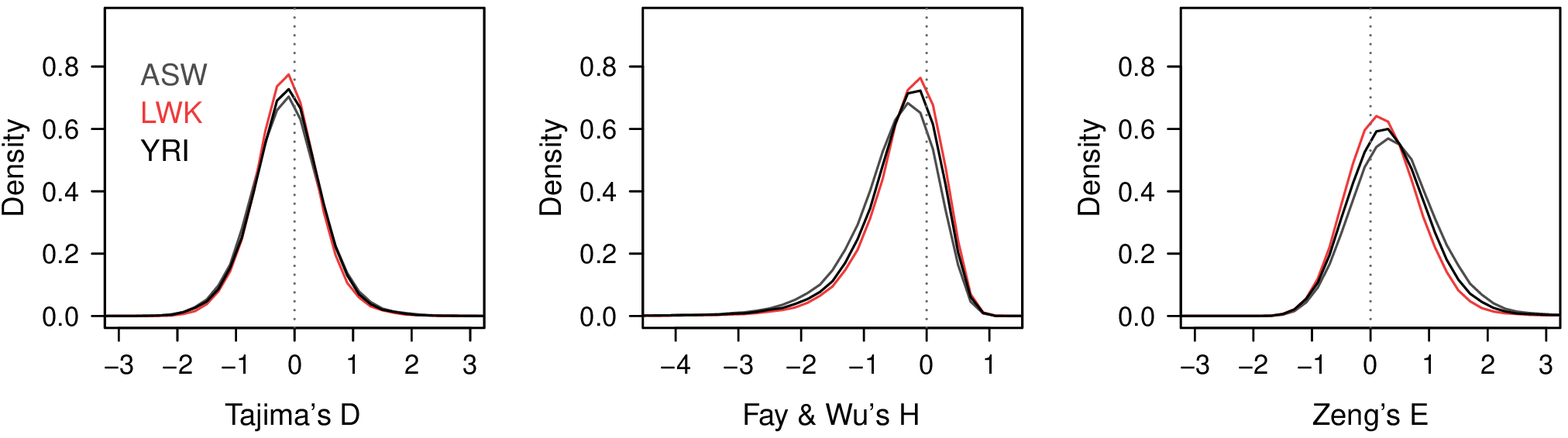}
\caption{\label{fig:densities_AFR}Distribution of test values over all sliding windows. Top row: original tests. 
Bottom row: demography-adjusted tests.}
\end{figure}
\begin{figure}[th]
\includegraphics[scale=0.8]{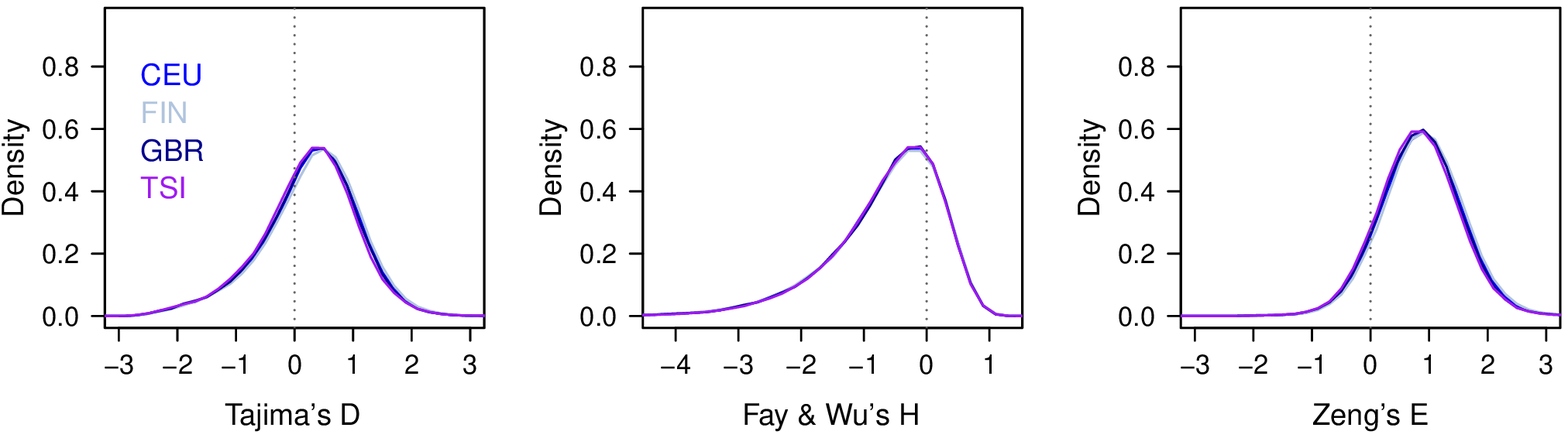}
\includegraphics[scale=0.8]{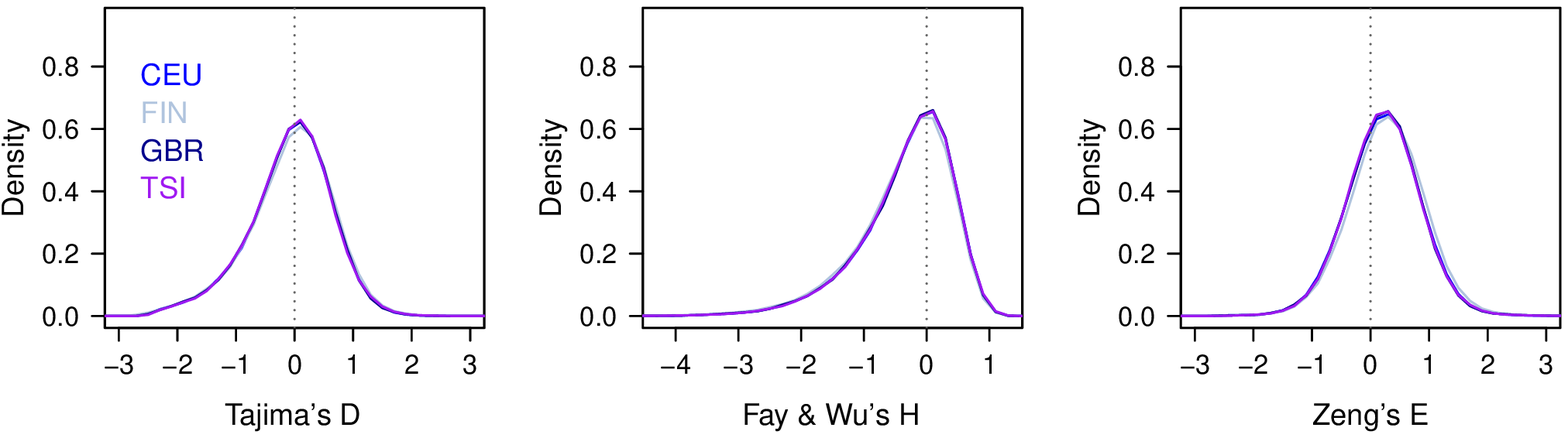}
\caption{\label{fig:densities_EUR}Distribution of test values over all sliding windows. Top row: original tests. 
Bottom row: demography-adjusted tests.}
\end{figure}
\begin{figure}[t]
\includegraphics[scale=0.8]{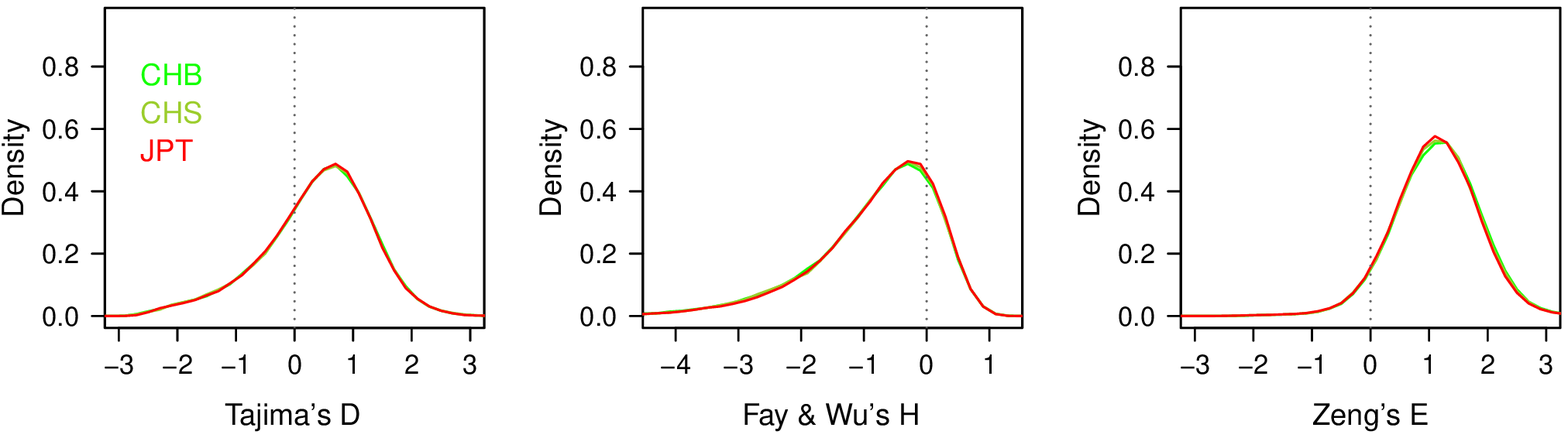}
\includegraphics[scale=0.8]{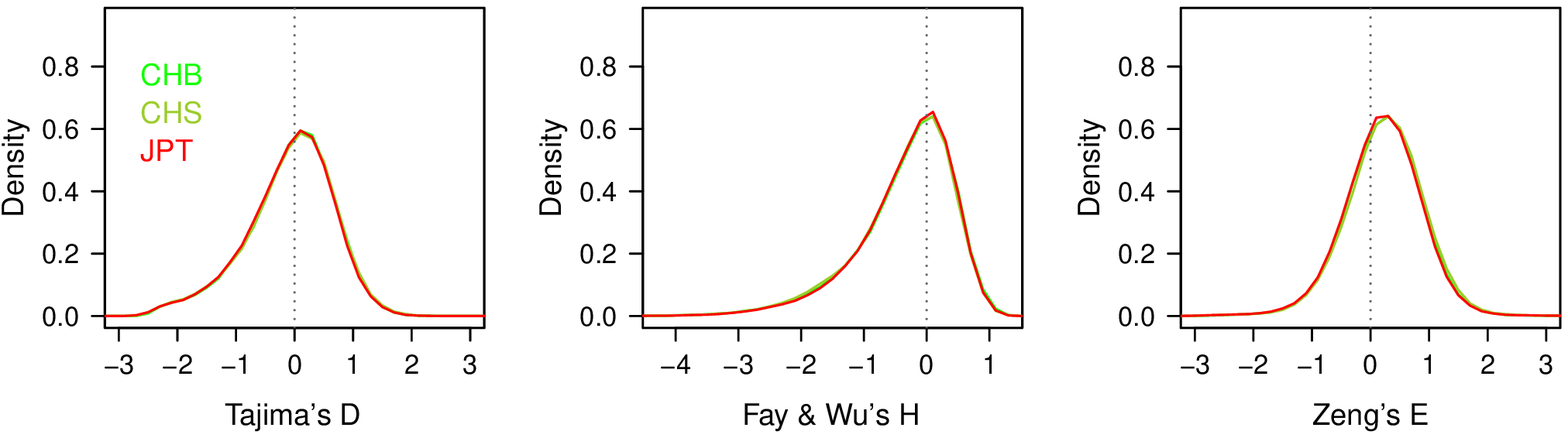}
\caption{\label{fig:densities_ASN}Distribution of test values over all sliding windows. Top row: original tests. 
Bottom row: demography-adjusted tests.}
\end{figure}
\newpage
\begin{sidewaystable}[ht]
\begin{flushleft}
\begin{small}
\begin{tabular}{rrrcl}
\multicolumn{3}{c}{Coordinates (hg19)} & Windows & Known genes (UCSC)\\
\hline
\multicolumn{5}{c}{ASW}\\
\hline
1 & 26.990.000 & 27.240.000 & 26 & ARID1A, PIGV, ZDHHC18, SFN, GPN2, GPATCH3, NR0B2, BC016143\\
2 & 95.560.000 & 95.790.000 & 24 & MAL, MRPS5\\
4 & 93.690.000 & 93.940.000 & 26 & GRID2\\
4 & 145.890.000 & 146.130.000 & 25 & ANAPC10, ABCE1, OTUD4, Mir$\_$649\\
5 & 45.000.000 & 45.280.000 & 29 & HCN1\\
5 & 133.980.000 & 134.190.000 & 22 & SEC24A, CAMLG, DDX46, C5orf24\\
16 & 14.620.000 & 14.810.000 & 20 & PARN, BFAR, PLA2G10, NPIP\\
16 & 46.470.000 & 46.660.000 & 20 & ANKRD26P1, SHCBP1\\
20 & 20.460.000 & 20.740.000 & 29 & \\
22 & 28.400.000 & 28.790.000 & 40 & Y$\_$RNA\\\hline
\multicolumn{5}{c}{LWK}\\\hline
1 & 41.500.000 & 41.710.000 & 22 &\\
2 & 95.560.000 & 95.760.000 & 21 & MAL, MRPS5\\
2 & 96.790.000 & 96.990.000 & 21 & DUSP2, CR749695, STARD7, LOC285033, TMEM127, CIAO1, SNRNP200\\
3 & 93.640.000 & 93.850.000 & 22 & ARL13B, STX19, DHFRL1, NSUN3, U7\\
8 & 99.600.000 & 99.930.000 & 34 &\\
11 & 66.390.000 & 66.600.000 & 22 & RBM14, RBM4, RBM4B, SPTBN2, C11orf80\\
17 & 44.210.000 & 44.400.000 & 20 & LOC644246, ARL17A, LRRC37A\\\hline
\multicolumn{5}{c}{YRI}\\\hline
1 & 41.500.000 & 41.720.000 & 23 &\\
2 & 95.560.000 & 95.810.000 & 26 & MAL, MRPS5\\
4 & 73.920.000 & 74.120.000 & 21 & COX18, ANKRD17\\
5 & 45.060.000 & 45.290.000 & 24 & HCN1\\
6 & 97.800.000 & 98.010.000 & 22 &\\
7 & 87.280.000 & 87.480.000 & 21 & RUNDC3B, SLC25A40\\
8 & 99.600.000 & 99.950.000 & 36 & 7SK\\
11 & 66.380.000 & 66.590.000 & 22 & RBM14, RBM4, RBM14-RBM4, RBM4B, SPTBN2, C11orf80\\
\end{tabular}
\caption{\label{tab:AFR} Contiguous regions of Tajima's D reduction (CRTR) in African populations.
}
\end{small}
\end{flushleft}
\end{sidewaystable}
\begin{sidewaystable}[ht]
\begin{flushleft}
\begin{small}
\begin{tabular}{rrrcl}
\multicolumn{3}{c}{Coordinates (hg19)} & Windows & Known genes (UCSC)\\
\hline
\multicolumn{5}{c}{CEU}\\
\hline
7 & 151.770.000 & 152.080.000 & 32 & GALNT11, MLL3\\
8 & 35.560.000 & 35.830.000 & 28 & UNC5D, AK092313\\
11 & 66.890.000 & 67.140.000 & 26 & KDM2A, DKFZp434M1735, ADRBK1, AK057681, ANKRD13D, SSH3, POLD4, 7SK, CLCF1, LOC100130987\\
15 & 44.240.000 & 44.440.000 & 21 & \\
15 & 44.580.000 & 44.890.000 & 32 & CASC4, CTDSPL2, LOC645212, EIF3J, SPG11\\
15 & 72.610.000 & 72.870.000 & 27 & HEXA, C15orf34, TMEM202, ARIH1\\
17 & 58.340.000 & 58.570.000 & 24 & C17orf64, L32131, APPBP2\\\hline
\multicolumn{5}{c}{FIN}\\\hline
1 & 35.680.000 & 36.120.000 & 45 & AF119915, ZMYM4, KIAA0319L, NCDN, TFAP2E, PSMB2\\
6 & 95.480.000 & 95.700.000 & 23 & \\
10 & 74.790.000 & 75.250.000 & 47 & NUDT13, BC069792, SNORA11, ECD, FAM149B1, DNAJC9, MRPS16, C10orf103, BC033983,
TTC18, ANXA7, \\
 & & & & ZMYND17, PPP3CB\\
12 & 89.020.000 & 89.230.000 & 22 & \\\hline
\multicolumn{5}{c}{GBR}\\\hline
1 & 27.930.000 & 28.140.000 & 22 & FGR, IFI6, FAM76A, STX12\\
1 & 35.680.000 & 36.110.000 & 44 & AF119915, ZMYM4, KIAA0319L, NCDN, TFAP2E, PSMB2\\
4 & 33.420.000 & 33.620.000 & 21 & \\
4 & 71.580.000 & 71.850.000 & 28 & RUFY3, GRSF1, MOB1B\\
8 & 35.580.000 & 35.830.000 & 26 & UNC5D, AK092313\\
11 & 66.890.000 & 67.140.000 & 26 & KDM2A, DKFZp434M1735, ADRBK1, AK057681, ANKRD13D, SSH3, POLD4, 7SK, CLCF1, LOC100130987\\
12 & 89.020.000 & 89.210.000 & 20 & \\
16 & 66.990.000 & 67.260.000 & 28 & CES3, CES4A, Metazoa$\_$SRP, CBFB, C16orf70, B3GNT9, BC007896, TRADD, FBXL8,
HSF4, NOL3,\\
 & & & &  KIAA0895L, EXOC3L1, E2F4, MIR328, ELMO3, LRRC29\\ 
17 & 58.490.000 & 58.770.000 & 29 & C17orf64,
L32131, APPBP2, PPM1D, BCAS3\\\hline\hline \multicolumn{5}{c}{TSI} \\\hline
1 & 35.690.000 & 36.110.000 & 43 & AF119915, ZMYM4, KIAA0319L, NCDN, TFAP2E, PSMB2\\
2 & 182.610.000 & 182.800.000 & 20 & SSFA2\\
8 & 35.600.000 & 35.850.000 & 26 & AK092313\\
8 & 42.720.000 & 43.000.000 & 29 & MIR4469, HOOK3, FNTA, SGK196, HGSNAT\\
10 & 75.130.000 & 75.350.000 & 23 & ANXA7, ZMYND17, PPP3CB, BC080555, USP54, U6\\
16 & 67.040.000 & 67.300.000 & 27 & Metazoa$\_$SRP, CBFB, C16orf70, B3GNT9, BC007896, TRADD, FBXL8, HSF4, NOL3,
KIAA0895L,\\
 & & & &  EXOC3L1, E2F4, MIR328, ELMO3, LRRC29, TMEM208, FHOD1, AK021876, SLC9A5\\ 
17 & 58.500.000 & 58.770.000 & 28 & L32131, APPBP2, PPM1D, BCAS3\\
\end{tabular}
\caption{\label{tab:EUR} Contiguous regions of Tajima's D reduction (CRTR) in European populations.}
\end{small}
\end{flushleft}
\end{sidewaystable}
\begin{sidewaystable}[ht]
\begin{flushleft}
\begin{small}
\begin{tabular}{ @{} rrrcl}
\multicolumn{3}{c}{Coordinates (hg19)} & Windows & Known genes (UCSC)\\
\hline
\multicolumn{5}{c}{CHB}\\
\hline
1 & 92.570.000 & 92.950.000 & 39 & KIAA1107, C1orf146, GLMN, RPAP2, GFI1\\
2 & 72.410.000 & 72.950.000 & 55 & U2, EXOC6B\\
2 & 108.980.000 & 109.550.000 & 58 & SULT1C4, GCC2, FLJ38668, LIMS1, RANBP2, CCDC138, EDAR\\
5 & 117.390.000 & 117.620.000 & 24 & BC044609\\
6 & 126.660.000 & 126.910.000 & 26 & CENPW, AK127472\\
11 & 60.920.000 & 61.140.000 & 23 & PGA3, PGA4, PGA5, VWCE, DDB1, DAK, CYBASC3, TMEM138\\
12 & 44.650.000 & 44.870.000 & 23 & \\
16 & 48.120.000 & 48.410.000 & 30 & ABCC12, ABCC11, LONP2, SIAH1, LOC100507577, MIR548AE2\\
16 & 67.220.000 & 67.580.000 & 37 & E2F4, MIR328, ELMO3, LRRC29, TMEM208, FHOD1, AK021876, SLC9A5, PLEKHG4, KCTD19,
LRRC36, U1, TPPP3,\\
 & & & & ZDHHC1, HSD11B2, ATP6V0D1, AGRP, FAM65A\\ 
20 & 30.190.000 & 30.390.000 & 21 & ID1, MIR3193,
COX4I2, BCL2L1, TPX2\\\hline 
\multicolumn{5}{c}{CHS}\\\hline
2 & 72.450.000 & 73.010.000 & 57 & U2, SNORD78, EXOC6B\\
3 & 17.340.000 & 17.860.000 & 53 & TRNA$\_$Pseudo\\
3 & 25.880.000 & 26.110.000 & 24 & LOC285326\\
5 & 117.380.000 & 117.620.000 & 25 & BC044609\\
8 & 67.500.000 & 68.140.000 & 65 & LOC645895, VCPIP1, C8orf44, PTTG3P, C8orf44-SGK3, SGK3, C8orf45, SNORD87, SNHG6,
TCF24, U2, PPP1R42,\\
 & & & & JA611241, COPS5, CSPP1, ARFGEF1\\ 
11 & 60.930.000 & 61.170.000 & 25 & PGA3, PGA4, PGA5, VWCE,
DDB1, DAK, CYBASC3, TMEM138, TMEM216\\ 
16 & 67.240.000 & 67.530.000 & 30 & LRRC29, TMEM208, FHOD1, AK021876, SLC9A5, PLEKHG4, KCTD19, LRRC36, U1, TPPP3,
ZDHHC1,\\
 & & & & HSD11B2, ATP6V0D1, AGRP\\\hline
 \multicolumn{5}{c}{JPT}\\\hline
1 & 87.350.000 & 87.540.000 & 20 & HS2ST1\\
2 & 72.410.000 & 73.080.000 & 68 & U2, SNORD78, EXOC6B\\
7 & 142.680.000 & 142.980.000 & 31 & OR9A2, OR6V1, OR6W1P, PIP, TAS2R39, TAS2R40, GSTK1\\
12 & 123.980.000 & 124.270.000 & 30 & MIR3908, TMED2, DDX55, EIF2B1, GTF2H3, TCTN2, ATP6V0A2, DNAH10\\
13 & 20.190.000 & 20.440.000 & 26 & MPHOSPH8, PSPC1, ZMYM5\\
16 & 48.110.000 & 48.380.000 & 28 & ABCC12, ABCC11, LONP2, MIR548AE2\\
16 & 67.230.000 & 67.590.000 & 37 & MIR328, ELMO3, LRRC29, TMEM208, FHOD1, AK021876, SLC9A5, PLEKHG4, KCTD19,
LRRC36, U1, TPPP3,\\
 & & & & ZDHHC1, HSD11B2, ATP6V0D1, AGRP, FAM65A\\
\end{tabular}
\caption{\label{tab:ASN} Contiguous regions of Tajima's D reduction (CRTR) in Asian populations.}
\end{small}
\end{flushleft}
\end{sidewaystable}
\newpage
\begin{sidewaystable}[ht]
\begin{flushleft}
\begin{small}
\begin{tabular}{rrrcl}
\multicolumn{3}{c}{Coordinates (hg19)} & Windows & Known genes (UCSC)\\
\hline
\multicolumn{5}{c}{ASW}\\
\hline
1 & 26.990.000 & 27.240.000 & 26 & ARID1A, PIGV, ZDHHC18, SFN, GPN2, GPATCH3, NR0B2, BC016143\\
2 & 95.560.000 & 95.760.000 & 21 & MAL, MRPS5\\
4 & 93.690.000 & 93.930.000 & 25 & GRID2\\
4 & 145.910.000 & 146.130.000 & 23 & ANAPC10, ABCE1, OTUD4, Mir$\_$649\\
5 & 45.060.000 & 45.280.000 & 23 & HCN1\\
16 & 46.470.000 & 46.660.000 & 20 & ANKRD26P1, SHCBP1\\
20 & 20.460.000 & 20.750.000 & 30 &\\
22 & 28.400.000 & 28.740.000 & 35 & Y$\_$RNA\\\hline
\multicolumn{5}{c}{LWK}\\\hline
1 & 41.500.000 & 41.710.000 & 22 & \\
3 & 93.670.000 & 93.860.000 & 20 & ARL13B, STX19, DHFRL1, NSUN3, U7\\
4 & 87.390.000 & 87.620.000 & 24 & PTPN13\\
8 & 99.600.000 & 99.930.000 & 34 & \\
11 & 66.390.000 & 66.590.000 & 21 & RBM14, RBM4, RBM4B, SPTBN2, C11orf80\\
12 & 87.490.000 & 87.680.000 & 20 & \\
17 & 44.210.000 & 44.400.000 & 20 & LOC644246, ARL17A, LRRC37A\\\hline
\multicolumn{5}{c}{YRI}\\\hline
1 & 41.490.000 & 41.720.000 & 24 & SCMH1\\
2 & 95.560.000 & 95.850.000 & 30 & MAL, MRPS5, ZNF514, ZNF2\\
5 & 45.070.000 & 45.290.000 & 23 & HCN1\\
6 & 97.800.000 & 97.990.000 & 20 & \\
8 & 99.600.000 & 99.950.000 & 36 & 7SK\\
11 & 66.380.000 & 66.620.000 & 25 & RBM14, RBM4, RBM14-RBM4, RBM4B, SPTBN2, C11orf80, RCE1, PC\\
\end{tabular}
\caption{\label{tab:AFRad} Contiguous regions of Tajima's D (demography-adjusted) reduction (CRTR) in African
populations}
\end{small}
\end{flushleft}
\end{sidewaystable}

\begin{sidewaystable}[ht]
\begin{flushleft}
\begin{small}
\begin{tabular}{rrrcl}
\multicolumn{3}{c}{Coordinates (hg19)} & Windows & Known genes (UCSC)\\
\hline
\multicolumn{5}{c}{CEU}\\\hline
1 & 35.720.000 & 35.920.000 & 21 & AF119915, ZMYM4, KIAA0319L\\
7 & 87.270.000 & 87.510.000 & 25 & RUNDC3B, SLC25A40, DBF4\\
7 & 151.770.000 & 152.080.000 & 32 & GALNT11, MLL3\\
8 & 35.570.000 & 35.840.000 & 28 & UNC5D, AK092313\\
11 & 66.880.000 & 67.140.000 & 27 & KDM2A, DKFZp434M1735, ADRBK1, AK057681, ANKRD13D, SSH3, POLD4, 7SK, CLCF1, LOC100130987\\
13 & 72.070.000 & 72.270.000 & 21 & \\
15 & 44.240.000 & 44.430.000 & 20 & \\
15 & 44.570.000 & 44.800.000 & 24 & CASC4, CTDSPL2\\
15 & 72.610.000 & 72.890.000 & 29 & HEXA, C15orf34, TMEM202, ARIH1, MIR630\\
17 & 58.340.000 & 58.570.000 & 24 & C17orf64, L32131, APPBP2\\\hline
\multicolumn{5}{c}{FIN}\\\hline
1 & 35.680.000 & 36.120.000 & 45 & AF119915, ZMYM4, KIAA0319L, NCDN, TFAP2E, PSMB2\\
3 & 96.470.000 & 96.660.000 & 20 & EPHA6\\
6 & 95.480.000 & 95.710.000 & 24 & \\
8 & 48.660.000 & 48.910.000 & 26 & PRKDC, MCM4\\
12 & 89.020.000 & 89.230.000 & 22\\
16 & 47.190.000 & 47.520.000 & 34 & Y$\_$RNA, ITFG1, PHKB\\\hline
\multicolumn{5}{c}{GBR}\\\hline
1 & 35.680.000 & 36.110.000 & 44 & AF119915, ZMYM4, KIAA0319L, NCDN, TFAP2E, PSMB2\\
4 & 33.420.000 & 33.610.000 & 20\\
6 & 128.440.000 & 128.650.000 & 22 & PTPRK\\
8 & 35.580.000 & 35.850.000 & 28 & UNC5D, AK092313\\
8 & 67.660.000 & 67.950.000 & 30 & PTTG3P, SGK3, C8orf45, SNORD87, SNHG6, TCF24, U2, PPP1R42\\
11 & 66.890.000 & 67.140.000 & 26 & KDM2A, DKFZp434M1735, ADRBK1, AK057681, ANKRD13D, SSH3, POLD4, 7SK, CLCF1, LOC100130987\\
16 & 66.970.000 & 67.260.000 & 30 & CES3, CES4A, Metazoa$\_$SRP, CBFB, C16orf70, B3GNT9, BC007896, TRADD, FBXL8,
HSF4, NOL3, KIAA0895L,\\
 & & & & EXOC3L1, E2F4, MIR328, ELMO3, LRRC29\\ 
 17 & 58.490.000 & 58.780.000 & 30 & C17orf64,L32131, APPBP2, PPM1D, BCAS3\\\hline 
\multicolumn{5}{c}{TSI}\\\hline
1 & 35.690.000 & 36.110.000 & 43 & AF119915, ZMYM4, KIAA0319L, NCDN, TFAP2E, PSMB2\\
4 & 33.430.000 & 33.620.000 & 20 &\\
8 & 35.570.000 & 35.860.000 & 30 & UNC5D, AK092313\\
8 & 42.720.000 & 43.000.000 & 29 & MIR4469, HOOK3, FNTA, SGK196, HGSNAT\\
16 & 67.040.000 & 67.310.000 & 28 & Metazoa$\_$SRP, CBFB, C16orf70, B3GNT9, BC007896, TRADD, FBXL8, HSF4, NOL3,
KIAA0895L,EXOC3L1,\\
 & & & &  E2F4, MIR328, ELMO3, LRRC29, TMEM208, FHOD1, AK021876, SLC9A5\\ 
 17 & 58.520.000 & 58.770.000 & 26) & APPBP2, PPM1D, BCAS3\\
\end{tabular}
\caption{\label{tab:EURad} Contiguous regions of Tajima's D (demography-adjusted) reduction (CRTR) in European
populations}
\end{small}
\end{flushleft}
\end{sidewaystable}

\begin{sidewaystable}[ht]
\begin{flushleft}
\begin{small}
\begin{tabular}{rrrcl}
\multicolumn{3}{c}{Coordinates (hg19)} & Windows & Known genes (UCSC)\\
\hline
\multicolumn{5}{c}{CHB}\\\hline
1 & 92.570.000 & 92.950.000 & 39 & KIAA1107, C1orf146, GLMN, RPAP2, GFI1\\
2 & 72.410.000 & 72.950.000 & 55 & U2, EXOC6B\\
2 & 108.980.000 & 109.440.000 & 47 & SULT1C4, GCC2, FLJ38668, LIMS1, RANBP2, CCDC138\\
5 & 117.390.000 & 117.620.000 & 24 & BC044609\\
6 & 126.660.000 & 127.030.000 & 38 & CENPW, AK127472, Vimentin3\\
11 & 60.920.000 & 61.150.000 & 24 & PGA3, PGA4, PGA5, VWCE, DDB1, DAK, CYBASC3, TMEM138\\
12 & 44.590.000 & 44.880.000 & 30 & \\
16 & 47.090.000 & 47.410.000 & 33 & NETO2, Y$\_$RNA, ITFG1\\
16 & 47.510.000 & 48.410.000 & 91 & PHKB, BC048130, ABCC12, ABCC11, LONP2, SIAH1, LOC100507577, MIR548AE2\\
16 & 67.190.000 & 67.850.000 & 67 & FBXL8, HSF4, NOL3, KIAA0895L, EXOC3L1, E2F4, MIR328, ELMO3, LRRC29, TMEM208,
FHOD1, AK021876,\\
 & & & & SLC9A5, PLEKHG4, KCTD19, LRRC36, U1, TPPP3, ZDHHC1, HSD11B2, ATP6V0D1, AGRP, FAM65A, CTCF\\
 & & & & DL491203,
RLTPR, ACD, PARD6A, C16orf48, C16orf86, AX747090, GFOD2, RANBP10, TSNAXIP1\\ 
20 & 30.120.000 & 30.370.000 & 26 & PSIMCT-1, HM13, ID1, MIR3193, COX4I2, BCL2L1, TPX2\\\hline 
\multicolumn{5}{c}{CHS}\\\hline
2 & 72.450.000 & 73.020.000 & 58 & U2, SNORD78, EXOC6B\\
2 & 82.540.000 & 82.810.000 & 28 & \\
3 & 17.350.000 & 17.830.000 & 49 & TRNA$\_$Pseudo\\
5 & 117.390.000 & 117.620.000 & 24 & BC044609\\
6 & 126.660.000 & 127.020.000 & 37 & CENPW, AK127472, Vimentin3\\
8 & 67.600.000 & 68.140.000 & 55 & PTTG3P, SGK3, C8orf45, SNORD87, SNHG6, TCF24, U2, PPP1R42, JA611241, COPS5, CSPP1, ARFGEF1\\
10 & 22.030.000 & 22.280.000 & 26 & DNAJC1, 7SK\\
11 & 60.930.000 & 61.200.000 & 28 & PGA3, PGA4, PGA5, VWCE, DDB1, DAK, CYBASC3, TMEM138, TMEM216, CPSF7, SDHAF2\\
12 & 88.480.000 & 88.760.000 & 29 & CEP290, TMTC3\\
16 & 47.080.000 & 47.410.000 & 34 & NETO2, Y$\_$RNA, ITFG1\\
16 & 47.430.000 & 48.140.000 & 72 & ITFG1, PHKB, BC048130, ABCC12\\
16 & 67.230.000 & 67.910.000 & 69 & MIR328, ELMO3, LRRC29, TMEM208, FHOD1, AK021876, SLC9A5, PLEKHG4, KCTD19,
LRRC36, U1, TPPP3,\\
& & & & ZDHHC1, HSD11B2, ATP6V0D1, AGRP, FAM65A, CTCF, DL491203, RLTPR, ACD, PARD6A, C16orf48,\\
& & & & C16orf86, AX747090, GFOD2, RANBP10, TSNAXIP1, CENPT, THAP11, NUTF2, EDC4\\\hline
\end{tabular}
\caption{\label{tab:CHad} Contiguous regions of Tajima's D (demography-adjusted) reduction (CRTR) in Chinese
populations}
\end{small}
\end{flushleft}
\end{sidewaystable}
\begin{sidewaystable}[ht]
\begin{flushleft}
\begin{small}
\begin{tabular}{rrrcl}
\multicolumn{3}{c}{Coordinates (hg19)} & Windows & Known genes (UCSC)\\
\hline
 \multicolumn{5}{c}{JPT}\\\hline 
1 &27.000.000 & 27.320.000 & 33 & ARID1A, PIGV, ZDHHC18, SFN, GPN2, GPATCH3, NR0B2, NUDC, C1orf172, BC016143\\ 
1 & 92.570.000 & 93.180.000 & 62 & KIAA1107, C1orf146, GLMN, RPAP2, GFI1, EVI5\\
2 & 72.410.000 & 73.080.000 & 68 & U2, SNORD78, EXOC6B\\
6 & 126.660.000 & 127.030.000 & 38 & CENPW, AK127472, Vimentin3\\
7 & 142.680.000 & 142.980.000 & 31 & OR9A2, OR6V1, OR6W1P, PIP, TAS2R39, TAS2R40, GSTK1\\
12 & 124.010.000 & 124.270.000 & 27 & MIR3908, TMED2, DDX55, EIF2B1, GTF2H3, TCTN2, ATP6V0A2, DNAH10\\
13 & 20.230.000 & 20.430.000 & 21 & PSPC1, ZMYM5\\
16 & 46.900.000 & 48.400.000 & 151 & GPT2, DNAJA2, NETO2, Y$\_$RNA, ITFG1, PHKB, BC048130, ABCC12, ABCC11, LONP2,
SIAH1,\\
 & & & & LOC100507577, MIR548AE2\\ 
16 & 67.180.000 & 67.750.000 & 58 & B3GNT9, BC007896, TRADD, FBXL8, HSF4, NOL3,
KIAA0895L, EXOC3L1, E2F4, MIR328, ELMO3, LRRC29,\\
 & & & & TMEM208, FHOD1, AK021876, SLC9A5, PLEKHG4, KCTD19, LRRC36, U1, TPPP3,
ZDHHC1, HSD11B2, \\
 & & & & ATP6V0D1, AGRP, FAM65A, CTCF, DL491203, RLTPR, ACD, PARD6A, C16orf48, C16orf86, AX747090, GFOD2\\
\end{tabular}
\caption{\label{tab:JPTad} Contiguous regions of Tajima's D (demography-adjusted) reduction (CRTR) in the Japanese
population}
\end{small}
\end{flushleft}
\end{sidewaystable}